\documentclass[conference]{IEEEtran}
\IEEEoverridecommandlockouts
\usepackage{cite}
\usepackage{amsmath,amssymb,amsfonts}
\usepackage[ruled,vlined]{algorithm2e}
\usepackage{graphicx}
\usepackage{textcomp}
\usepackage{xcolor}
\usepackage{soul}
\usepackage{subfig}
\usepackage{tabularx}
\usepackage{comment}
\usepackage{svg}
\usepackage{multirow}
\usepackage{enumitem}
\usepackage[utf8]{inputenc}
\usepackage{setspace}
\usepackage{hyperref}
\usepackage{cleveref}

\def\BibTeX{{\rm B\kern-.05em{\sc i\kern-.025em b}\kern-.08em
    T\kern-.1667em\lower.7ex\hbox{E}\kern-.125emX}}
\definecolor{burntorange}{rgb}{0.8, 0.33, 0.0}
\definecolor{darkblue}{rgb}{0.0, 0.0, 0.55}
\definecolor{deepcarrotorange}{rgb}{0.91, 0.41, 0.17}
\definecolor{darkgreen}{rgb}{0.0, 0.2, 0.13}
\definecolor{darkred}{rgb}{0.55, 0.0, 0.0}
\SetCommentSty{mycommfont}

\begin{document}

\title{Investigating Misinformation in Online Marketplaces: An Audit Study on Amazon}

\author{\IEEEauthorblockN{Eslam Hussein}
\IEEEauthorblockA{Department of Computer Science \\
Virginia Tech\\
Email: ehussein@vt.edu}
\and
\IEEEauthorblockN{Hoda Eldardiry}
\IEEEauthorblockA{Department of Computer Science \\
Virginia Tech\\
Email: hdardiry@vt.edu}
}

\maketitle
\thispagestyle{plain}
\pagestyle{plain}

\setstretch{0.87}

\begin{abstract}
Search and recommendation systems are ubiquitous and irreplaceable tools in our daily lives. Despite their critical role in selecting and ranking the most relevant information, they typically do not consider the veracity of information presented to the user. In this paper, we introduce an audit methodology to investigate the extent of misinformation presented in search results and recommendations on online marketplaces. We investigate the factors and personalization attributes that influence the amount of misinformation in searches and recommendations. Recently, several media reports criticized Amazon for hosting and recommending items that promote misinformation on topics such as vaccines. Motivated by those reports, we apply our algorithmic auditing methodology on Amazon to verify those claims. Our audit study investigates (a) factors that might influence the search algorithms of Amazon and (b) personalization attributes that contribute to amplifying the amount of misinformation recommended to users in their search results and recommendations. Our audit study collected $\sim$526k search results and $\sim$182k homepage recommendations, with $\sim$8.5k unique items. Each item is annotated for its stance on vaccines’ misinformation (pro, neutral or anti). Our study reveals that (1) the selection and ranking by the default \emph{Featured} search algorithm of search results that have misinformation stances are positively correlated with the stance of search queries and customers’ evaluation of items (ratings and reviews), (2) misinformation stances of search results are neither affected by users’ activities nor by interacting (browsing, wish-listing, shopping) with items that have a misinformation stance, and (3) a filter bubble built-in users’ homepages have a misinformation stance positively correlated with the misinformation stance of items that a user interacts with.

\end{abstract}
%\linespread{0.95}
\section{Introduction}
People depend on search and recommendation systems to look for information and make decisions without understanding how the algorithms that operate these systems work~\cite{10.1145/2702123.2702556}.
Recently, Amazon -- one of the largest online marketplaces -- has been critiqued by several mainstream media for hosting and recommending items that deliberately present misleading information about vaccination~\cite{article1, article4}.
Vaccines' misinformation significantly increases the rates of vaccine hesitancy~\cite{misinfoHesitancy}, which is declared by the World Health Organization (WHO) as one of the main 10 threats to global health in 2019~\cite{who3}. Personalization is an integral part of almost every search and recommendation system, where presented information is tailored to users based on personalization attributes (e.g., users’ demographics and user-system history).

In this work, we present a methodology to study the prevalence of algorithmically curated misinformative search results and recommendations in online marketplaces. This is done by systematically examining the effect of personalization on the extent of misinformation presentation. We examine how the misinformation stance (pro, neutral and anti) of items being \emph{browsed}, \emph{added to wish list} and \emph{added to cart} would affect the stance of personalized search results and recommendations. In particular, we focus on auditing \emph{Amazon's} search and recommendation algorithms, in order to understand how items are presented in search results and recommendations with respect to their stance toward vaccines' misinformation. 
We also investigate whether the platform presents misinformative items about vaccines on search and recommendations, and the key factors that might influence the underlying algorithms.

\textbf{The key contributions of this work can be summarized as follows:} 1) We present a methodology to study the prevalence of algorithmically curated misinformative search results and recommendations in online marketplaces. This study is the first to systematically examine the effect of personalization on the extent of misinformation returned in search results and recommendations on online marketplaces. 2) We contribute to the research community by building a rich dataset consisting of $\sim$526k search results, $\sim$182k homepage recommendations and 8566 unique items annotated for their stances toward vaccines' misinformation, along with the personalization attributes audited~\footnote{Data will be published in an online public repository upon acceptance along with a ReadMe file that describes the dataset in detail.}. 3) Our study revealed how personalization leads to building a filter bubble of recommendations in a user's homepage. 4) Our study investigates Amazon's default \emph{Featured} search algorithm by analyzing what factors drive the selection and ranking of items in search results given the items' misinformation stance.

\section{Research Questions}\label{rq:RQs}
The main objectives of this work is to investigate how misinformative items get ranked and recommended to a user in search results and recommendations, and to understand what personalization attributes that contribute into amplifying the amount of misinformation in search results and recommendations. To achieve those objectives we guide our research to answer the following three research questions. Note that, we focus on Amazon as the online marketplace platform and on vaccines as the misinformation topic throughout this work.\\
\noindent\textbf{RQ1 - [SEARCH ALGORITHMS]. Do search algorithms steer users toward more misinformative search results? What are the contributing factors?}\label{rq:RQ1} To investigate this question, we examine three major factors that might influence the search algorithms;
\begin{itemize}
    \item \textbf{RQ1a - Ranking Algorithm.}\label{rq:RQ1a} How items with  misinformation stance get ranked by the 5 search algorithms?
    \item \textbf{RQ1b - Search Query.}\label{rq:RQ1b} Is there a correlation between misinformation stances of search queries and search results? i.e., would pro-misinformative search queries (e.g., "vaccine illusion") generate more misinformative search results than neutral search queries (e.g., "vaccines")?
    \item \textbf{RQ1c - User Rating.}\label{rq:RQ1c} How the ratings of an item by users affect its rankings in search results, given that the item has a stance toward misinformation?
\end{itemize}

\noindent\textbf{RQ2 - [USER ACTIVITY]. What is the effect of a user activity on the amount of misinformative search results and recommendations?}\label{rq:RQ2}
\begin{itemize}
\item \textbf{RQ2-H.} We hypothesize that the closer the user to purchase a specific item the more similar items would be present in search results and recommendations. Hence adding misinformative items to a shopping cart would generate more misinformative search results and recommendations than adding the same items to the user's wish list. Similarly, adding misinformative items to a wish list would generate more misinformative search results and recommendations than just browsing these items.
\end{itemize}

\noindent\textbf{RQ3 - [ITEM STANCE]. Given user history (browsed, wished for or purchased) for an item, what is the effect of the item stance toward misinformation on the amount of generated misinformed search results and recommendations?}\label{rq:RQ3}
\begin{itemize}
    \item \textbf{RQ3-H.} We hypothesize that whenever a user interacts with an item of a particular misinformation stance (pro misinformation, neutral or anti misinformation), the search results and recommendation would have a similar misinformation stance.
\end{itemize}

\section{Background and Related Work}

%%%%%%%%%%%%%%%%%%%%%%%%%%%%%%%%%%%%%%%%%%%%%%%%%%%%%%%%%%%%%%%%%%%%
%%%%%%   Misinformation in search and recommendation systems
%%%%%%%%%%%%%%%%%%%%%%%%%%%%%%%%%%%%%%%%%%%%%%%%%%%%%%%%%%%%%%%%%%%%
\textbf{Misinformation in search results and recommendations.}
Misinformation has been widely studied under various themes including conspiracy theories~\cite{10.1371/journal.pone.0118093}, rumors~\cite{ICWSM148122, 10.1145/3341161.3342916}, hoaxes~\cite{10.1145/2872427.2883085}, fake news~\cite{nature1, Lazer1094, Vosoughi1146} and information credibility~\cite{ICWSM1510582, 10.1145/1963405.1963500}. Most research studied misinformation in social media~\cite{10.1145/3161603, 7396773, 10.1257/jep.31.2.211, 10.1145/3305260, ICWSM1817907, shibutani1966improvised}.
Even though $\sim$92\% of the U.S. population acquires information through search and recommendation systems~\cite{purcell2011findings}, very limited research methodically inspects the algorithmic curation of misinformative content on such systems~\cite{10.1145/3392854, doi:10.1080/10810730.2020.1776423}.
Most online users are unaware that search results and recommendations are algorithmically curated and personalized. Some users do not know that search and recommendation systems might present inaccurate information. One study revealed that $\sim$62\% of Facebook users were unaware of Facebook News Feed recommendation algorithm~\cite{10.1145/2702123.2702556}. Another study discovered that a significant number of YouTube users came to believe that Earth is flat after they watched recommended videos~\cite{articleconsp}. Another study found that searching for “vitamin k shot” on YouTube and Google returned results discouraging taking the vitamin shot; and some of the top 10 results from YouTube and Google were recommending anti-vaccine content~\cite{article0}. Another study examining politically related bias in YouTube videos showed that videos supporting Trump promoted conspiracies and fake news about Clinton~\cite{article6}. Prior work also found that content consumed through search is usually highly trusted and can change people’s political views and voting behaviors~\cite{druckman2005impact}. Algorithmically curated and presented content has been investigated due to concerns that algorithmically-targeted content may have influenced the 2016 U.S. election~\cite{article5}. Search and recommendation systems sometimes recommend and present biased and misinformative content~\cite{10.1145/3159652.3159732}. 

%%%%%%%%%%%%%%%%%%%%%%%%%%%%%%%%%%%%%%%%%%%%%%%%%%%%%%%%%%%%%%%%%%%%
%%%%%%   Algorithmic audit studies
%%%%%%%%%%%%%%%%%%%%%%%%%%%%%%%%%%%%%%%%%%%%%%%%%%%%%%%%%%%%%%%%%%%%

\begin{figure}
\centering
  \includegraphics[scale=0.11]{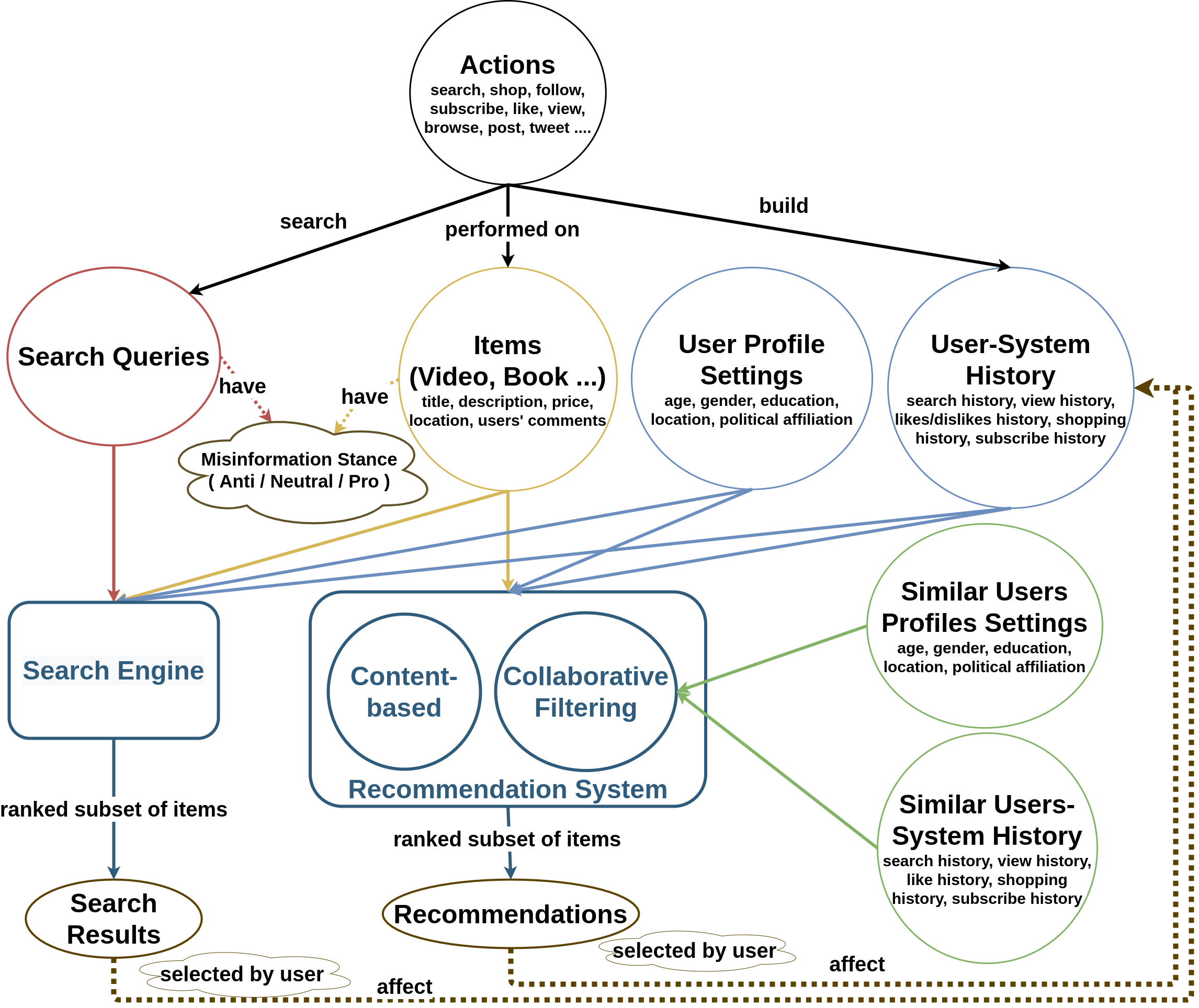}
  \caption{\scriptsize Search and recommendation system components and their interdependencies.}
  \label{fig:search_recommednation_systems}
\end{figure}

\textbf{Search and recommendation systems.}
Search engines and recommendation systems are integral components of the ecosystems of many online platforms. Algorithms that operate these components aim to perform two key tasks: (1) search and (2) recommend tailored content to the user. In search, the user provides a search query and the search algorithm selects, ranks and returns a list of relevant search results. For selection and ranking, the algorithm utilizes additional attributes besides the search query to tailor the search results. Additional attributes include user demographic information (e.g., gender, age, location) and user system history (e.g., previous searches, browsing history, purchase history). Recommendation algorithms usually employ one or a hybrid of popular recommendation models~\cite{ISINKAYE2015261, ISI:000352789600020, ISI:000401965900004} including content-based and collaborative filtering models.
Content-based recommendation algorithms tailor recommendations based on similarities (e.g., titles, descriptions, ratings, prices) between items being recommended and items previously selected, liked or purchased by the user. User demographics are also considered. Collaborative filtering algorithms are similar to content-based algorithms, but in addition, consider attitudes and preferences of similar users (i.e., users who share similar demographics, system-history or preferences). Recommendation algorithms aim to curate a set of recommendations that increase user satisfaction, sales, and subscriptions to the platform; thereby increasing the overall platform revenue. In Fig.~\ref{fig:search_recommednation_systems}, we illustrate a graphical representation of search and recommendation system components and their interdependencies.

\textbf{Auditing search and recommendation algorithms.}
Due to the opacity of search and recommendation algorithms, they have been commonly investigated as a black-box using algorithmic auditing techniques. This involves systematic statistical analysis of the algorithms of an online platform to detect suspicious behavior~\cite{sandvig2014auditing}. These methods include repeatedly querying algorithms and observing the corresponding output. This provides an understanding of algorithms internal mechanics without detailed knowledge of its internal process. Algorithmic auditing has proved its efficacy in studying (a) bias in online systems \cite{hannak2014measuring, chen2016empirical, 10.1145/3173574.3174225, 10.1145/2702123.2702520, 10.1145/2998181.2998327}, (b) the extent of personalization used by search engines~\cite{10.1145/2488388.2488435, 10.1145/2815675.2815714}, and (c) the extent of online misinformation~\cite{10.1145/3392854, doi:10.1080/10810730.2020.1776423}.

Previous work attempted to investigate the prevalence of vaccines' misinformation on Amazon, but the study has some limitations~\cite{doi:10.1080/10810730.2020.1776423}. First, search was performed using only a single query. Our study presented in this paper uses 29 search queries which enables the exploration of several dimensions within Amazon's search space. Another limitation of the previous study is the limited focus on a single item type; namely, books. Moreover, the study only considered 104 books, which is a relatively small dataset. On the other hand, our dataset is much larger (8566 unique items) and covers several types of items (e.g., books, videos, audibles, etc.). Another major limitation of the previous study is that it does not investigate any personalization effect on search and recommendation. However, our study investigates various personalization attributes. The last limitation, the previous study investigates only one search algorithm (\emph{Featured}), while our study investigates all five existing Amazon algorithms.

\textbf{Vaccines' misinformation and hesitancy.}
Vaccination is one of the most successful methods of preventing the spread of infectious diseases.
Yet recently, increasing numbers of parents doubt vaccines' efficacy and safety fearing possible side effects on children. Vaccine controversies are based on misinformed beliefs that vaccines contain deleterious ingredients such as Mercury and Aluminum that can lead to diseases such as autism and sudden infant death syndrome. Some conspiracy theories claim that the propaganda behind mandating vaccination is promoted by big pharmaceutical companies to increase their revenues. Others claim that some diseases can be cured alone by the human body’s immune system and therefore vaccination is not required. Such misinformation is refuted by WHO, other authoritative agencies, and scientific studies~\cite{who1, who2, cdc1, NAP10997, TAYLOR20143623}.
Such conspiracies and pseudoscience-based claims on vaccines lead to increasing rates of vaccine hesitancy~\cite{misinfoHesitancy}.

\section{Methodology}
\begin{figure}[t]
  \centering
  \vspace{-3mm}
  \hspace{-3mm}
  \subfloat[]{{\label{fig:amz_queries}\includegraphics[scale=0.2]{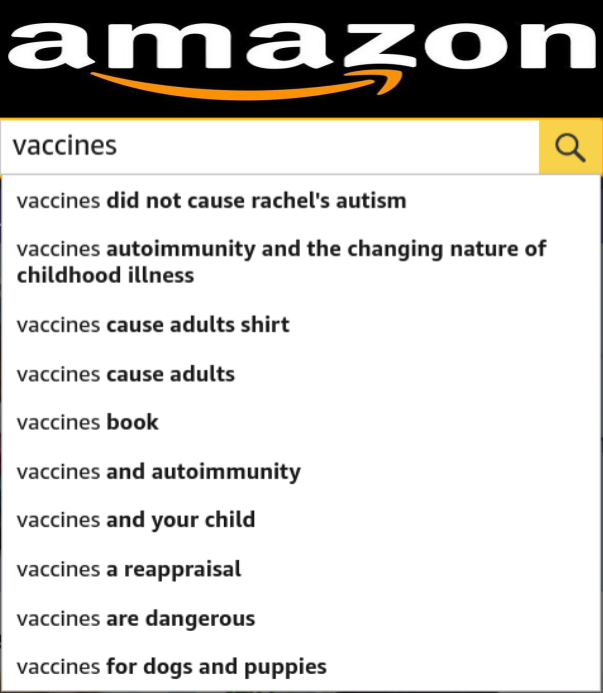} }}\hfill
  \subfloat[]{\label{fig:gt_queries}{\includegraphics[scale=0.2]{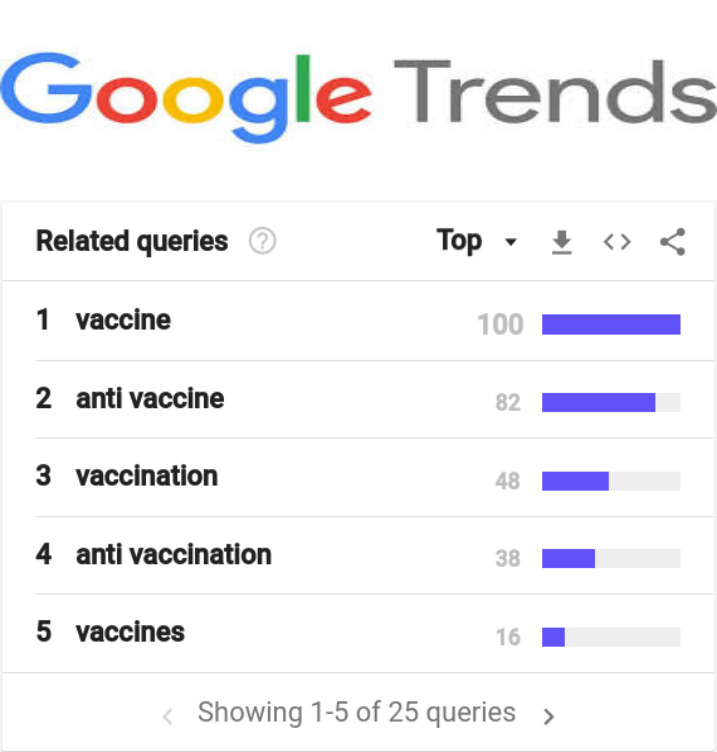}}}
%   \vspace{-10pt}
\vspace{-3mm}
  \caption{\scriptsize (a) List of Amazon’s auto-complete suggestions when the seed query “vaccines” fed into the search box at the top of Amazon’s page, (b) Top related queries used to search in the Vaccine Controversies topic in Google Trends.}
  \label{fig:search_queries}
   \vspace{-8pt}
\end{figure}

\textbf{Search queries.} One of the key challenges in conducting audit studies is the selection of search queries that are a) related to the topic under investigation, b) most used on the platform and c) most recently used to search for that topic. We build on our recently published query selection methodology~\cite{10.1145/3392854} to identify high impact search queries related to the topic of vaccine controversies. Our goal is to curate a set of search queries that satisfies two requirements: (1) include search queries that are most used to search for information about vaccines controversies, and (2) include search queries that are most recently used by users on Amazon. To meet these requirements we collect queries from two sources: a) Google Trends; where we generate analytics about the topic of "Vaccine Controversies" (see fig. \ref{fig:gt_queries}) and get a list of the top related queries that we include to our query list. b)- Amazon search box (see fig.~\ref{fig:amz_queries}); where we feed seed queries (vaccine(s), vaccination) and collect a list of auto-complete query suggestions generated by Amazon. This list represents the most recently used queries by Amazon users. We add those queries to our query list. Later, we follow a set of heuristics to shortlist our query list in order to remove redundant queries, semantically similar queries (e.g., “vaccines” and “vaccine”) and longer queries (length$>$4 words) that are overtly specific (e.g., “vaccines did not cause Rachel’s autism”). Our final query set consists of 29 queries that represent the most high impact search queries. We annotate each of the 29 search queries for the stance of its expected search results toward vaccines' misinformation, for example, query "vaccine illusion" is annotated with 1 (pro misinfo) since the expected search results for that query would be vaccines' misinformative. Another example, query "vaccine" would be annotated with 0 (neutral) since the expected search results would be neutral to vaccines' misinformation. Check the queries curation and annotation algorithm \ref{algo:search_queries}.

\begin{algorithm}
\scriptsize
\SetAlgoLined
\KwResult{query\_set; Set of most related, used and recent annotated queries}
% = \{\}\;
\tcp{top related queries by Google Trends}
query\_set = GoogleTrends.TopQueries(topic=Vaccine Controversies)\;
\tcp{most recently used queries in Amazon}
query\_set += AmazonSearchBox(seeds=[vaccine(s), vaccination])\;
\tcp{follow heuristics to shortlist query set}
query\_set -= \{redundant queries, semantically similar queries, longer queries, overtly specific queries\}\;

\ForEach{query $q \in query\_set$}{
  \uIf{expected\_results(q) are pro-misinfo}{
   annotate(q, +1)\;
   }
   \uElseIf{expected\_results(q) are anti-misinfo}{
   annotate(q, -1)\;
   }
   \Else{
   annotate(q, 0)\;
  }
}
 \caption{Search queries curation algorithm}
 \label{algo:search_queries}
\end{algorithm}

\textbf{Annotation.} We ran our audit experiment repeatedly for 14 consecutive days and collected $\sim$526k items in search results including 6870 unique items, and $\sim$182k recommendations from homepages including 1762 unique items. Together, search and recommendation data include a total of 8566 unique items. To develop proper annotation heuristics for our datasets, we randomly sampled 100 items and manually annotated them by three different individuals separately. After iterations of discussions, we developed the following annotation heuristics and classes:
\begin{itemize}
    \item Items related to vaccination are given values of -1, 0 or 1, where items that promote vaccines' misinformation or discourage vaccination are annotated with 1, items that oppose, debunk or mock vaccines' misinformation or encourage vaccination are given -1, and items that are neutral will be given a value of 0.
    \item  Items not related to the topic of vaccination will be give a value of 2, items presented in a foreign language than English will be given 3, and items that were removed from Amazon after data collection will be given a value of 4. 
\end{itemize}
Next to agreeing on the annotation classes and heuristics, one person annotated all of the 8566 unique items. To quantify the amount of misinformation, annotation values given to each class are normalized. The goal of normalizing the annotation stances is to calculate the amount of misinformation on a Search Engine Results Page (SERP) or a homepage on a scale from -1 (all items against misinformation) to +1 (all items are pro misinformation). This informs the decision on whether a personalization attribute leads to more or less misinformative search results or recommendations. This fine grained 6-point scale system is then mapped into a 3-point scale system (-1, 0 and 1). Annotation values -1, 0 and 1 remain the same after the normalization. Annotation value 2 is mapped to 0, and items annotated as 3 or 4 are ignored. Table~\ref{tab:annotation} shows the description, heuristics, normalized scores, counts and examples of each annotation class.

\textbf{Misinformation Score of a SERP.} We measure the extent of misinformation in each of the saved SERPs using the SERP Misinformation Score (SERP-MS)~\cite{10.1145/3392854}\footnote{SERP-MS = $\frac{\sum_{r=1}^{n} {(x_r * (n - r+1))} }{ \frac{n * (n + 1)}{2}}$; where $r$ is the rank of the search result and $n$ is the number of search results present in the SERP}, which computes the amount of misinformation in a SERP while considering the SERP ranking of results. A SERP-MS ranges from -1 (all search results oppose misinformation) to +1 (all search results promote misinformation).

\textbf{Misinformation Score of a Recommendation page.} In Amazon, a user's homepage is composed of items recommended inside components. Each component contains a set of items that are both (a) ranked horizontally and (b) belong to the same recommendation heuristic (e.g., "Related to items you've viewed", "Inspired by your shopping trends"). A SERP composed of results combined horizontally and vertically is called a federated SERP \cite{10.1145/1935826.1935922}. We develop a new score based on SERP-MS called \textbf{FSERP-MS} (Federated SERP Misinformation Score) that does a similar job by measuring the misinformation in a page by considering both (a) the rank of the component in the page and (b) the rank of the item within the component. Equation~\ref{eqn:fserp} shows the FSERP-MS formula; where $r$ is the rank of a recommendation within a component, $n$ is the number of recommendations inside a component, $i$ is the rank of a component inside the federated SERP and $m$ is the number of components in a federated SERP. FSERP-MS is a continuous value ranging from -1 (all items in all components are oppose misinformation) to +1 (all items in all components promote misinformation).

\begin{equation}\label{eqn:fserp}
\scriptsize
FSERP-MS = \frac{\sum_{i=1}^{m} [{(\frac{\sum_{r=1}^{n} {(x_r * (n - r+1))} }{ \frac{n * (n + 1)}{2}}) * (m - i+1)}] }{ \frac{m * (m + 1)}{2}}
\end{equation}

%\noindent
\begin{scriptsize}
\begin{table}
\setlength\tabcolsep{3pt}
\scriptsize
\centering
\begin{tabular}{|l|c|c|c|c|c|}
\hline
\multicolumn{1}{|c|}{{\color[HTML]{333333} }}                                    & {\color[HTML]{333333} }                                                                                     & \multicolumn{4}{c|}{{\color[HTML]{333333} \textbf{Stance Treatment}}}                                                                                                      \\ \cline{3-6} 
\multicolumn{1}{|c|}{\multirow{-2}{*}{{\color[HTML]{333333} \textbf{Activity}}}} & \multirow{-2}{*}{{\color[HTML]{333333} \textbf{\begin{tabular}[c]{@{}c@{}}No. of\\ accounts\end{tabular}}}} & {\color[HTML]{333333} \textbf{Pro Misinfo}} & {\color[HTML]{333333} \textbf{Neutral}} & {\color[HTML]{333333} \textbf{Anti Misinfo}} & {\color[HTML]{333333} \textbf{Mix}} \\ \hline
{\color[HTML]{333333} \textbf{Browse}}                                           & {\color[HTML]{333333} 4}                                                                                    & {\color[HTML]{333333} 1}                    & {\color[HTML]{333333} 1}                & {\color[HTML]{333333} 1}                     & {\color[HTML]{333333} 1}            \\ \hline
{\color[HTML]{333333} \textbf{Add to wish list}}                                 & {\color[HTML]{333333} 4}                                                                                    & {\color[HTML]{333333} 1}                    & {\color[HTML]{333333} 1}                & {\color[HTML]{333333} 1}                     & {\color[HTML]{333333} 1}            \\ \hline
\textbf{Add to cart}                                                             & 4                                                                                                           & 1                                           & 1                                       & 1                                            & 1                                   \\ \hline
\textbf{Search}                                                                  & \multicolumn{5}{c|}{1}                                                                                                                                                                                                                                                                   \\ \hline
\end{tabular}
\caption{\scriptsize 13 accounts created to execute the audit experiment, where 12 accounts will perform the 3 activities (browse, add to list and add to cart) on the 4 stance treatments (3X4) and 1 account created to solely search without having history at 11am along with the other 12 accounts.}
\label{tab:accounts}
\vspace{-8pt}
\end{table}
\end{scriptsize}

\begin{scriptsize}
\begin{table*}
\setlength\tabcolsep{1pt}
\scriptsize
\centering
\begin{tabular}{|c|l|l|c|c|l|}
\hline
\textbf{\begin{tabular}[c]{@{}c@{}}Annota-\\ tion Value\end{tabular}} & \multicolumn{1}{c|}{\textbf{Description}}                                                                                      & \multicolumn{1}{c|}{\textbf{Annotation Heuristics}}                                                                                                                                                                         & \textbf{\begin{tabular}[c]{@{}c@{}}No. of\\ unique items\end{tabular}}   & \textbf{\begin{tabular}[c]{@{}c@{}}Normali-\\ zed Score\end{tabular}} & \multicolumn{1}{c|}{\textbf{\begin{tabular}[c]{@{}c@{}}Sample items\\ (Title, URL {[}amazon.com/dp/asin{]})\end{tabular}}} \\ \hline
-1                                                                    & \begin{tabular}[c]{@{}l@{}}opposing, debunking or mocking\\ misinformation about vaccines\\ or promoting vaccines\end{tabular} & \begin{tabular}[c]{@{}l@{}}(1) refutes, disapprove or mocks misinformation about vaccines\\ (2) gives authoritative evidence that proves vaccines' safety or \\ efficacy (3) has a positive stance on vaccines\end{tabular} & \begin{tabular}[c]{@{}c@{}}search (524)\\ recommend (25)\end{tabular}    & -1                                                                & \begin{tabular}[c]{@{}l@{}}Vaccines: What Every Parent\\ Should Know, asin=0028638611\end{tabular}                         \\ \hline
0                                                                     & neutral                                                                                                                        & \begin{tabular}[c]{@{}l@{}}Item that does not have a positive or negative stance toward\\ vaccines\end{tabular}                                                                                                            & \begin{tabular}[c]{@{}c@{}}search (433)\\ recommend (13)\end{tabular}    & 0                                                                 & \begin{tabular}[c]{@{}l@{}}Vaccines and Autoimmunity,\\ asin=B014T25FX6\end{tabular}                                       \\ \hline
1                                                                     & \begin{tabular}[c]{@{}l@{}}promoting, supporting or\\ explaining misinformation\\  about vaccines\end{tabular}                 & \begin{tabular}[c]{@{}l@{}}(1) promotes, supports or adduces misinformation on vaccines\\ (2) promotes delaying or stopping vaccination or has a negative\\ stance on vaccines\end{tabular}                                             & \begin{tabular}[c]{@{}c@{}}search (419)\\ recommend (60)\end{tabular}    & 1                                                                 & \begin{tabular}[c]{@{}l@{}}A Shot in the Dark,\\ asin=089529463X\end{tabular}                                              \\ \hline
2                                                                     & Not about vaccines                                                                                                             & Item that is not related to vaccination                                                                                                                                                                                     & \begin{tabular}[c]{@{}c@{}}search (5464)\\ recommend (1658)\end{tabular} & 0                                                                 & \begin{tabular}[c]{@{}l@{}}PlayStation 4 Slim 1TB Console,\\ asin=B071CV8CG2\end{tabular}                                  \\ \hline
3                                                                     & Non English                                                                                                                    & \begin{tabular}[c]{@{}l@{}}Item that has its title, description or contents\\ in a non-English language\end{tabular}                                                                                                        & \begin{tabular}[c]{@{}c@{}}search (25)\\ recommend (0)\end{tabular}      & ignored                                                               & \begin{tabular}[c]{@{}l@{}}TETYANA OBUKHANYCH -\\ I VACCINI, asin=8893195445\end{tabular}                                  \\ \hline
4                                                                     & Removed                                                                                                                        & \begin{tabular}[c]{@{}l@{}}Item removed from Amazon at the time of\\ annotation\end{tabular}                                                                                                                                & \begin{tabular}[c]{@{}c@{}}search (5)\\ recommend (6)\end{tabular}       & ignored                                                               & \begin{tabular}[c]{@{}l@{}}Indica Plateau Don't Confuse\\ Your Search ..., asin=B07ZMJNJQ7\end{tabular}                    \\ \hline
\end{tabular}
\caption{\scriptsize Description and heuristics of each annotation class along with no. of unique items in search results and recommendations, normalized score and a sample item.}
\label{tab:annotation}
\end{table*}
\end{scriptsize}

\section{Experimental Design}
\begin{figure}[t]
  \centering
  \subfloat[\scriptsize Amazon SERP]{{\label{fig:amz_serp}\includegraphics[scale=0.2]{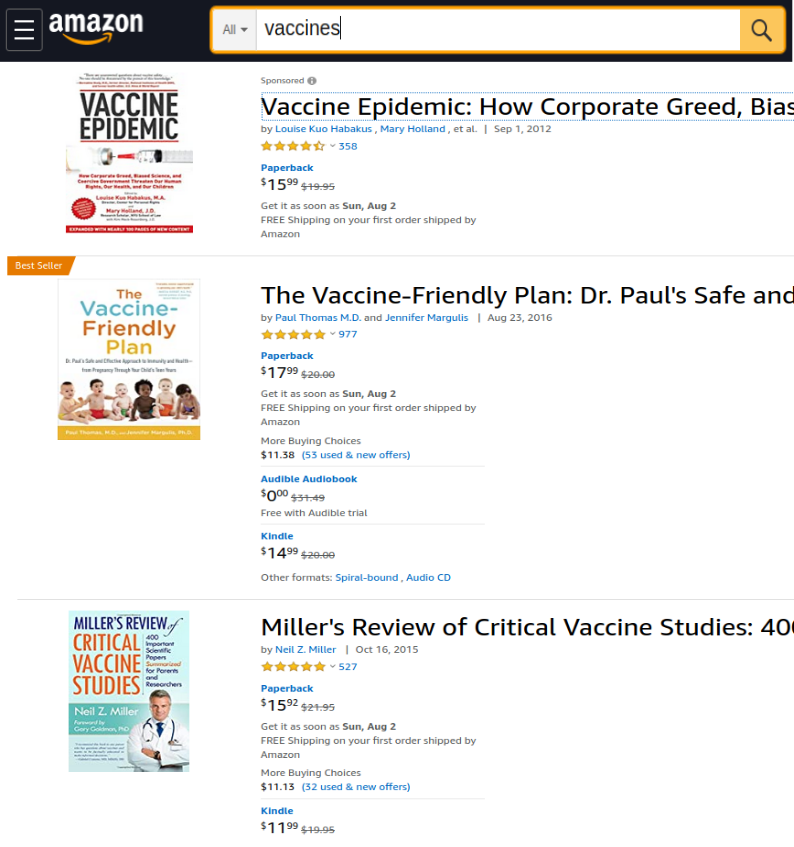} }}\hfill
  \subfloat[\scriptsize Amazon Homepage]{\label{fig:amz_recommendations}{\includegraphics[scale=0.2]{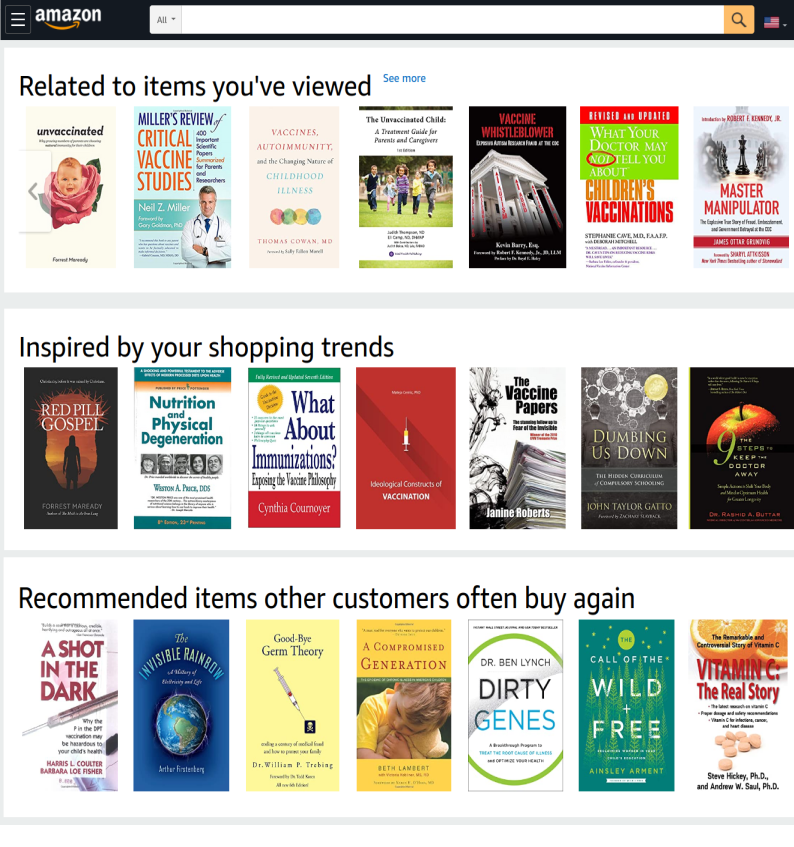}}}
   \vspace{-5pt}
  \caption{\scriptsize (a) An Amazon SERP generated when searching for “vaccines”, (b) An Amazon homepage of three recommendation components.}
  \label{fig:serp_recommendations}
%   \vspace{-10pt}
\end{figure}

We demonstrate our experimental design and execution to answer the research questions mentioned in sec.~\ref{rq:RQs}. In our experiments we collect search results and recommendations from two different Amazon pages: (a) User’s \emph{homepage}: We collect the first 20 recommendations from each recommendation component that exists on the user’s homepage on Amazon, fig.~\ref{fig:amz_recommendations} show three components on a homepage, where each component has a rank on the homepage and recommending no. of items, (b) Amazon \emph{SERP}: we collect the first 20 search results recommended by Amazon’s search algorithm when a user searches for a query, fig~\ref{fig:amz_serp} shows an example of an Amazon SERP.

\textbf{Measured personalization attributes.} We design our audit experiment to measure the amount of personalized misinformative search results and recommendations generated by Amazon. We audit the effect of personalization caused by the following two attributes:
\begin{itemize}
    \item Activities performed by user on Amazon, where we study the effect of four main activities: 1)- \emph{search}: user searches Amazon for the 29 queries 2)- \emph{browse}: user opens the page of an item and browses its content, 3)- \emph{add to wish  list}: user browses the page of an item and adds that item to his/her wish list by clicking on the “Add to list” button and 4)- \emph{add to cart}: user browses the page of an item and adds it to his/her shopping cart by clicking  on the “Add to cart” button. All activities and histories that each activity builds with the system are shown in table \ref{tab:actions_histories}.
    \item Stance of an item toward vaccines' misinformation (promote, neutral and oppose misinformation about vaccines). First we search each of the 29 search queries and collect the first 20 search results from their respective SERPs, some queries return fewer results (358 unique results in total), then we annotate each item in the set of unique search results with regard to its stance toward vaccines' misinformation. Next, we select the top 12 most rated items under each stance into three sets that represent the three stances. From those three sets, we create 4 items stance treatments to measure the effect of the item’s stance as shown in table \ref{tab:treatments}. We end up having four different stance treatments: \emph{Pro} treatment that contains 12 most rated items that have misinformative stance toward vaccines.  \emph{Neutral} treatment that contains 12 most rated items that have a neutral stance toward vaccines. \emph{Anti} treatment that contains 12 most rated items that oppose vaccines' misinformation or promote vaccines. And finally \emph{Mix} treatment that contains 4 from each of the 3 other treatments, we mix those 12 items randomly. Refer to table~\ref{tab:annotation} for heuristics and samples of the three main stances (with annotation values of -1, 0 or 1).
\end{itemize}

\textbf{Controlled personalization attributes.} We control for the personalization effects stemming form user's demographics (age, gender and location), because Amazon do not enable their users to set their age or gender during the sign-up process or through their Amazon accounts settings. Also, Amazon provides different methods to set the location of a user, for example an Amazon account could have different shipping addresses, and users also can set their locations through their accounts settings, in addition to the geographic location inferred from a user's IP address. That's why we control for the user's location by (1) executing our audit experiment for each Amazon account from the same location (Mountain View, CA), (2) configuring the location of each Amazon account to Mountain View, CA and (3) not adding any shipment address to any account.

\begin{scriptsize}
\begin{table}
%    \vspace{-10pt}
    \scriptsize
    \centering
    \begin{tabular}{|l|l|}
    \hline
    \multicolumn{1}{|c|}{\textbf{Activity}} & \multicolumn{1}{c|}{\textbf{User History}} \\ \hline
    Search                                & Search history                                  \\
    Browse                                & Browsing history                                  \\
    Add to wish list                      & Browsing history + Wish history                   \\
    Add to cart                           & Browsing history + Purchase history               \\ \hline
    \end{tabular}
    \vspace{-2pt}
    \caption{\scriptsize List of activities performed by Amazon accounts and history built by each user action.}
    \label{tab:actions_histories}
\end{table}
\end{scriptsize}

\begin{scriptsize}
\begin{table}
\setlength\tabcolsep{2pt}
\scriptsize
\centering
\begin{tabular}{|l|l|}
\hline
\multicolumn{1}{|c|}{\textbf{Treatment name}} & \multicolumn{1}{c|}{\textbf{Selection Criteria}}                                                                         \\ \hline
Anti Misinformation                           & 12 most rated items opposing vaccines' misinformation                                                                  \\ \hline
Neutral                                       & 12 most rated items that are neutral toward vaccines                                                                 \\ \hline
Pro Misinformation                            & 12 most rated items promoting vaccines' misinformation                                                                 \\ \hline
Mix                                           & \begin{tabular}[c]{@{}l@{}}12 items randomly mixed from the top 4 most rated items\\ from each stance (4 pro + 4 neutral + 4 anti)\end{tabular} \\ \hline
\end{tabular}
    \vspace{-2pt}
    \caption{\scriptsize List of items stance treatments used to measure the effect of an item’s stance toward vaccines' misinformation (Pro, Neutral or Anti)}
    \label{tab:treatments}
\vspace{-10pt}
\end{table}
\end{scriptsize}

\textbf{Controlling for noise.} In audit studies, noise could significantly affect search results and recommendations. For example, temporal noise attributed to regular updates of search indices could affect the returned search results if not controlled.
During the execution of our audit experiment, we control for different sources of noise, for example we control for browser noise by selecting the same version of Mozilla Firefox for all accounts where cookies are enabled and every browser history is cleaned daily before the execution of the experiment. We also control for temporal effect by performing all activities and searches for all accounts simultaneously. In addition, we control for the machines used in the experiment by having every machine configured similarly (Ubuntu 14.04, same generation of CPU and 3.75GB RAM), controlling for the machine configuration assures that no noise results from using different speeds of CPUs, different sizes of memory or varying performance due to different Operating Systems. Carry-over noise happens when a search operation affects the search results of the next search in two successive searches. Previous audit studies on Google Search showed that carry-over noise is noticed if the time between two successive searches is less than 11 minutes \cite{10.1145/2488388.2488435}. We use this as the benchmark and decide to keep a time interval of 20 minutes between two successive searches to control for noise from carry-over effects.

%%%% query vs results graphs
\begin{scriptsize}
\begin{figure}[t]
    \centering
    \vspace{-5mm}
    \includegraphics[scale=0.22]{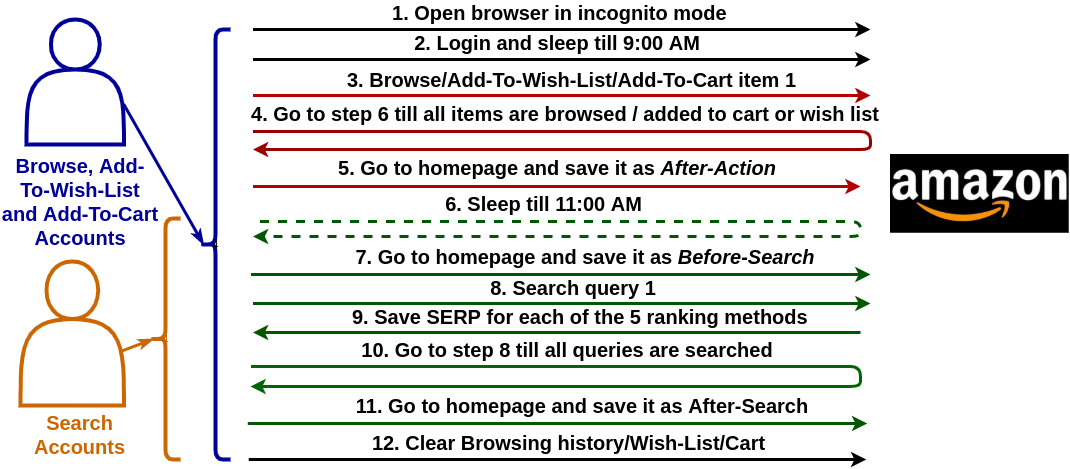}
    \caption{\scriptsize Steps taken by accounts that perform the four activities (\textcolor{deepcarrotorange}{Search}, \textcolor{darkblue}{Browse}, \textcolor{darkblue}{add to wish list} and \textcolor{darkblue}{add to cart}).}
    \label{fig:activities}
   \vspace{-15pt}
\end{figure}
\end{scriptsize}

\textbf{Auditing steps.} In order to measure the effects of user’s activities and item’s misinformation stance on Amazon search results and recommendation we designed the audit experiment as follows; we create a total of 13 Amazon accounts that are automated by Selenium scripts as follows ; 12 Amazon accounts where each will perform one of the last three activities shown in table \ref{tab:actions_histories} on one of the stance treatments shown in table \ref{tab:treatments}. We also create 1 Amazon account to solely search Amazon without having any history (browsing/wishing/purchasing) with the platform. The accounts layout of the experiment is shown in table \ref{tab:accounts}. For 14 consecutive days at 9am each of the accounts that will browse, add items to wish list or cart will start performing its assigned activity on each item in its assigned treatment set. After finishing all their respective activities, every account will save the homepage in order to measure the effect of activities and treatments on the recommendations generated on the homepage (After-Action), then all accounts pause till 11am. At 11 AM all of the 13 accounts - including the one that solely search - will save their homepage (Before-Search) and start searching for each of the 29 queries simultaneously. After searching each query the account will save the generated SERP of that query and pause for 20 minute in order to neutralize any carry-over effect. After searching for all queries, each bot goes to the account homepage and saves it (After-Search). Fig. \ref{fig:activities} shows steps taken by each account responsible for one of the four activities.

\section{Results}
In this section we analyze the data collected during our audit experiment to answer our research questions listed in section~\mbox{\ref{rq:RQs}}. We test our data for normality and we find that it is not normally distributed, also our samples have unequal sizes. Thus, we choose non-parametric tests. For pairwise comparisons, we use Mann-Whitney U test. For multiple comparisons, we use Kruskal-Wallis ANOVA followed by Tukey HSD for post-hoc analysis. Note that the SERP-MS score is computed for each SERP, while FSERP-MS score is computed for each homepage.

\subsection{RQ1:Search Algorithms\label{res:RQ1}}
In this section, we investigate how items that have a stance (Pro/Neutral/Anti) toward vaccines' misinformation get ranked SERPs and what factors might influence their presence and ranking. First, we investigate how each of the five ranking algorithms of Amazon would rank search results with respect to their stance toward misinformation. Second, we inspect the effect of a query stance to vaccines misinformation on the misinformation stance of search results. Last, we examine the user ratings of items under each misinformation stance to understand its influence on the prevalence of misinformative search results.

\subsubsection{RQ1a: Ranking Algorithms}\label{results:RQ1a}
We compare the rankings of items under each misinformation stance across the five search algorithms. We find that items under each stance rank significantly different across the five algorithms (\textit{Kruskal Wallis H(4)$\approx$51.6, p$\approx$1.7e-10}), Tukey-HSD reveals that both \emph{Featured} and \emph{Average Customer Review} algorithms rank misinformative items similarly without significant difference, while the other three algorithm also have a similar rankings for misinformative items, see fig. \ref{fig:rankings}. The same hold for neutral items with (\textit{Kruskal Wallis H(4)$\approx$52.18, p$\approx$1.3e-10}). On the other hand, items that are against vaccines misinformation rank significantly different across all algorithms (\textit{Kruskal Wallis H(4)$\approx$32, p$\approx$1.9e-6}), where $\mu$\textsubscript{Avg. Cust. Review} $<$ $\mu$\textsubscript{featured}, $\mu$\textsubscript{avg. customer review} $<$ $\mu$\textsubscript{price asc.} and $\mu$\textsubscript{price desc.} $<$ $\mu$\textsubscript{price asc.}. In summary, we deduce that the average customers reviews play a critical role in ranking items by the default (\emph{Featured}) search algorithm, therefore we find the rankings of both neutral and pro misinformative items are similar across both algorithms.

%%%% rankings graphs
\begin{scriptsize}
\begin{figure}[t]
  \centering
  \vspace{-4mm}
  %\hspace{-10pt}
  \subfloat[\scriptsize Featured]{{ \label{fig:ranking_featured_ranks}\includegraphics[scale=0.3]{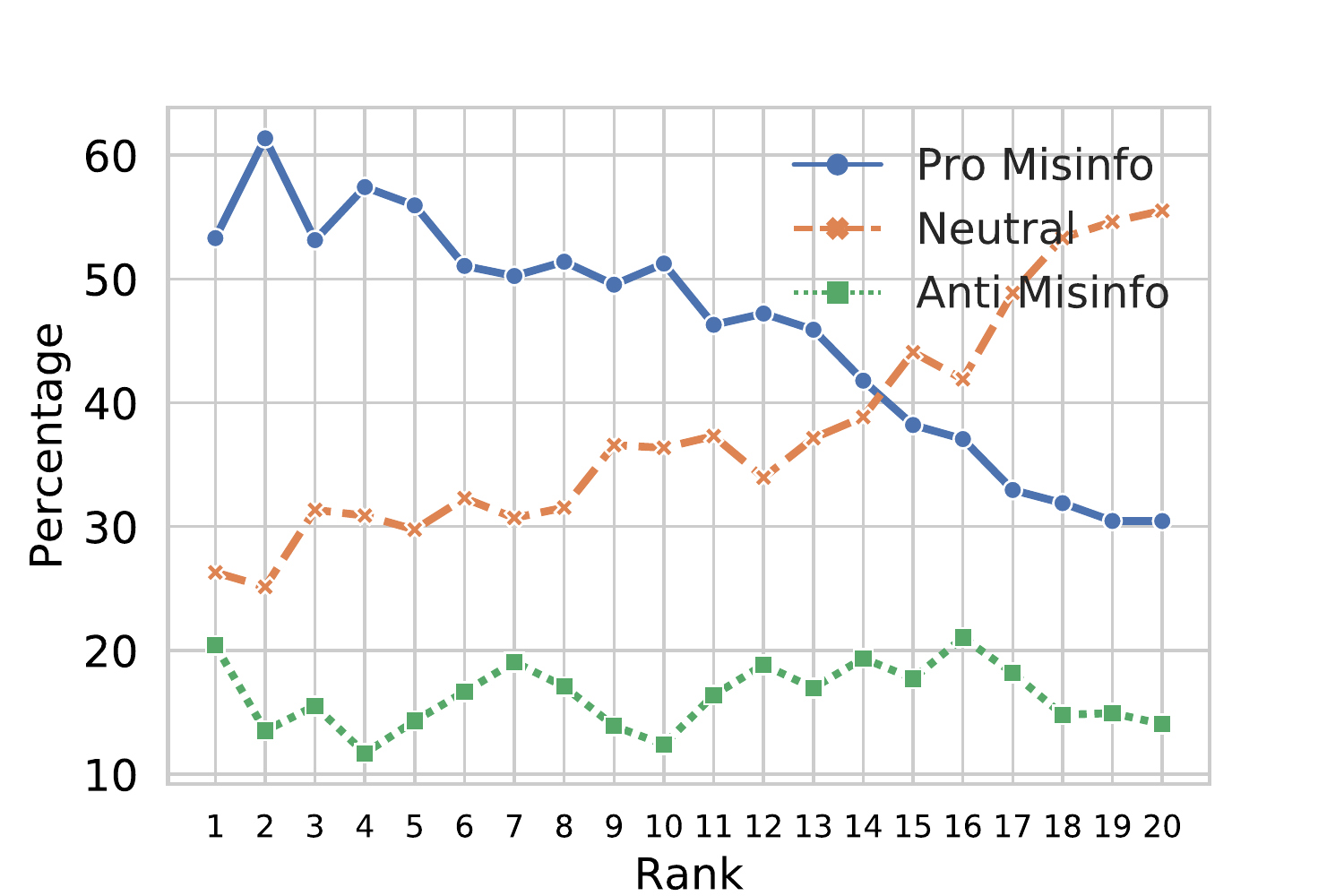} }}
  \subfloat[\scriptsize Average Customer Review]{{ \label{fig:ranking_avg_reviews_ranks}\includegraphics[scale=0.3]{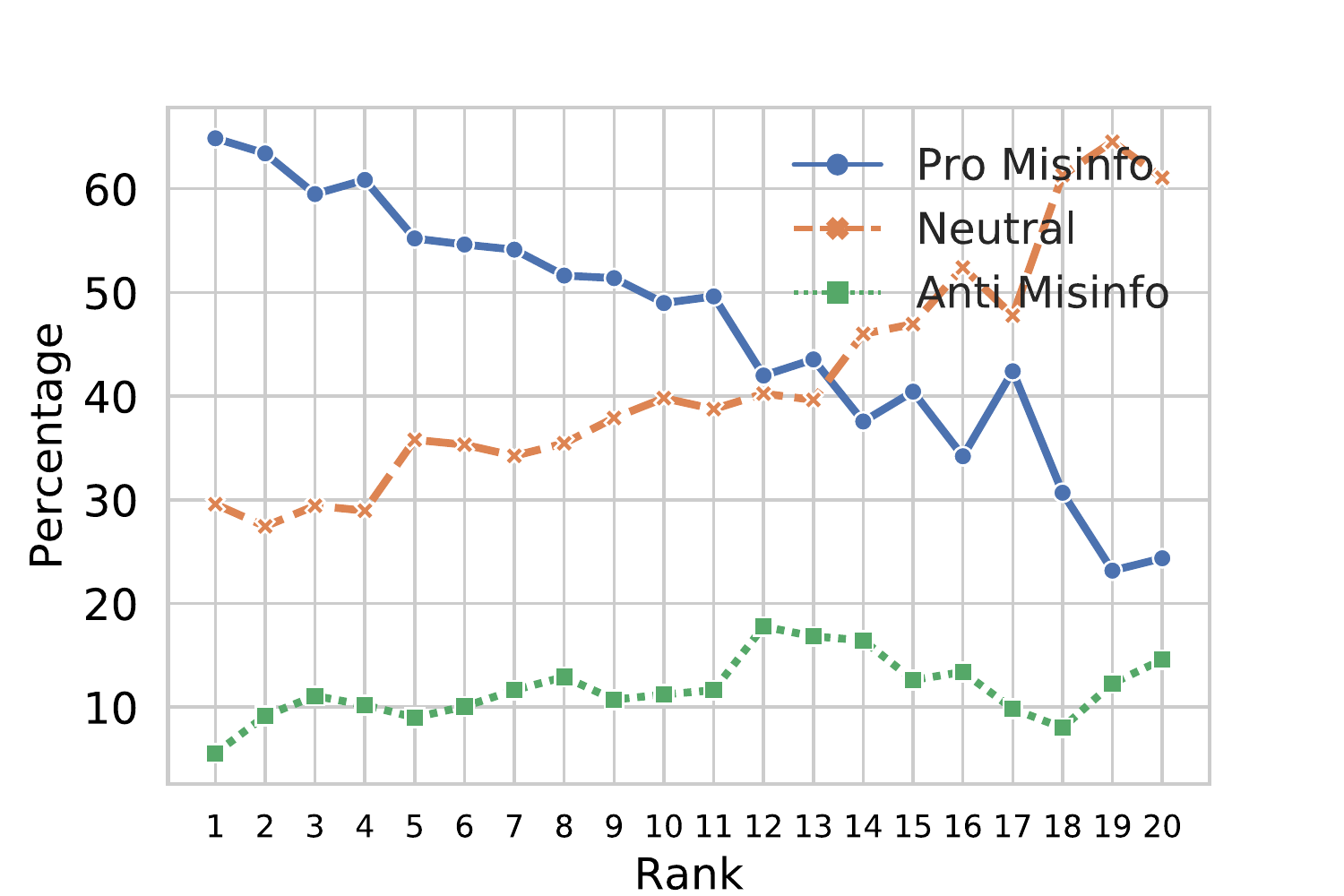} }}\\
  \subfloat[\scriptsize Price (Ascending)]{{ \label{fig:ranking_price_asc_ranks}\includegraphics[scale=0.3]{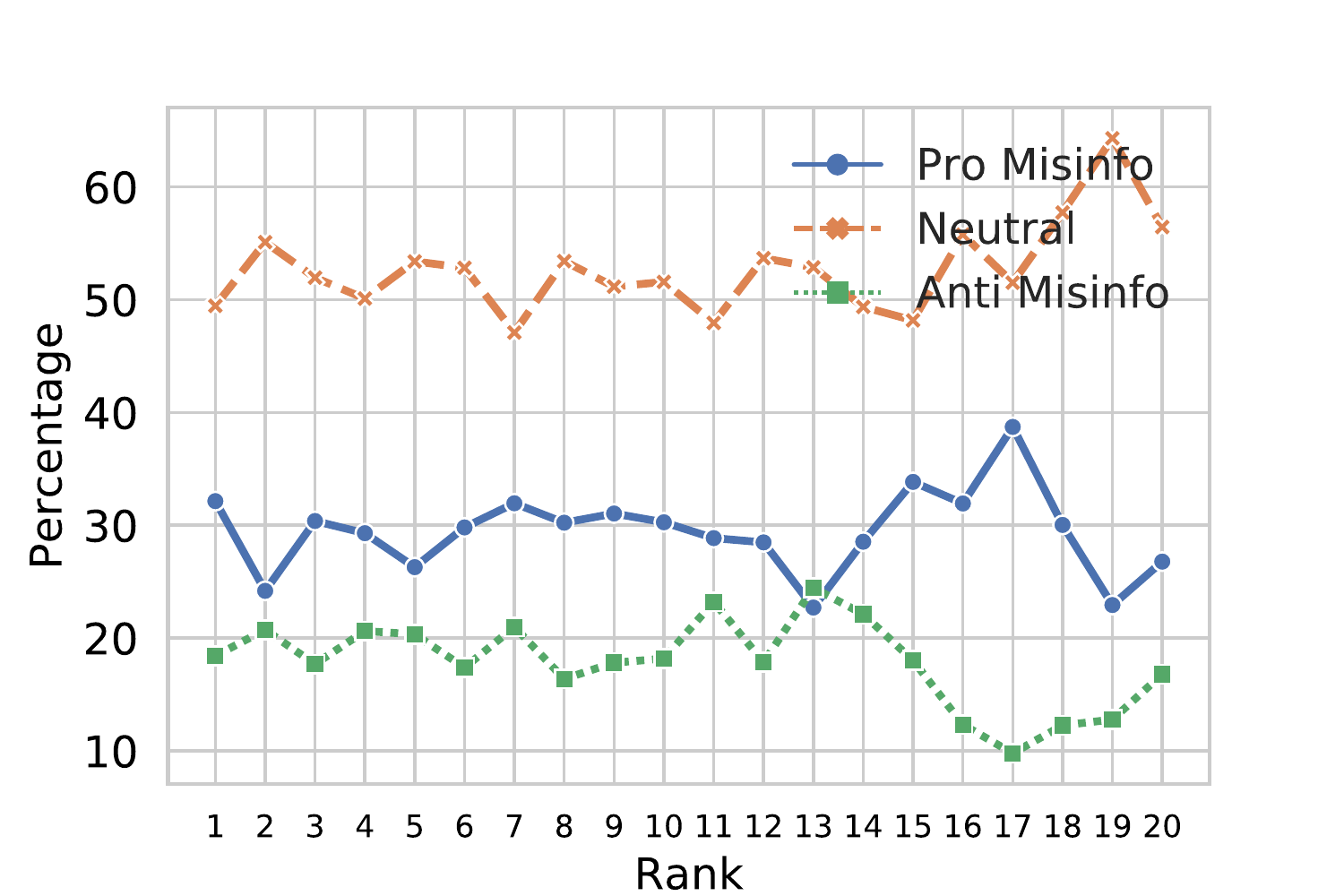} }}
  \subfloat[\scriptsize Price (Descending)]{{ \label{fig:ranking_price_dsc_ranks}\includegraphics[scale=0.3]{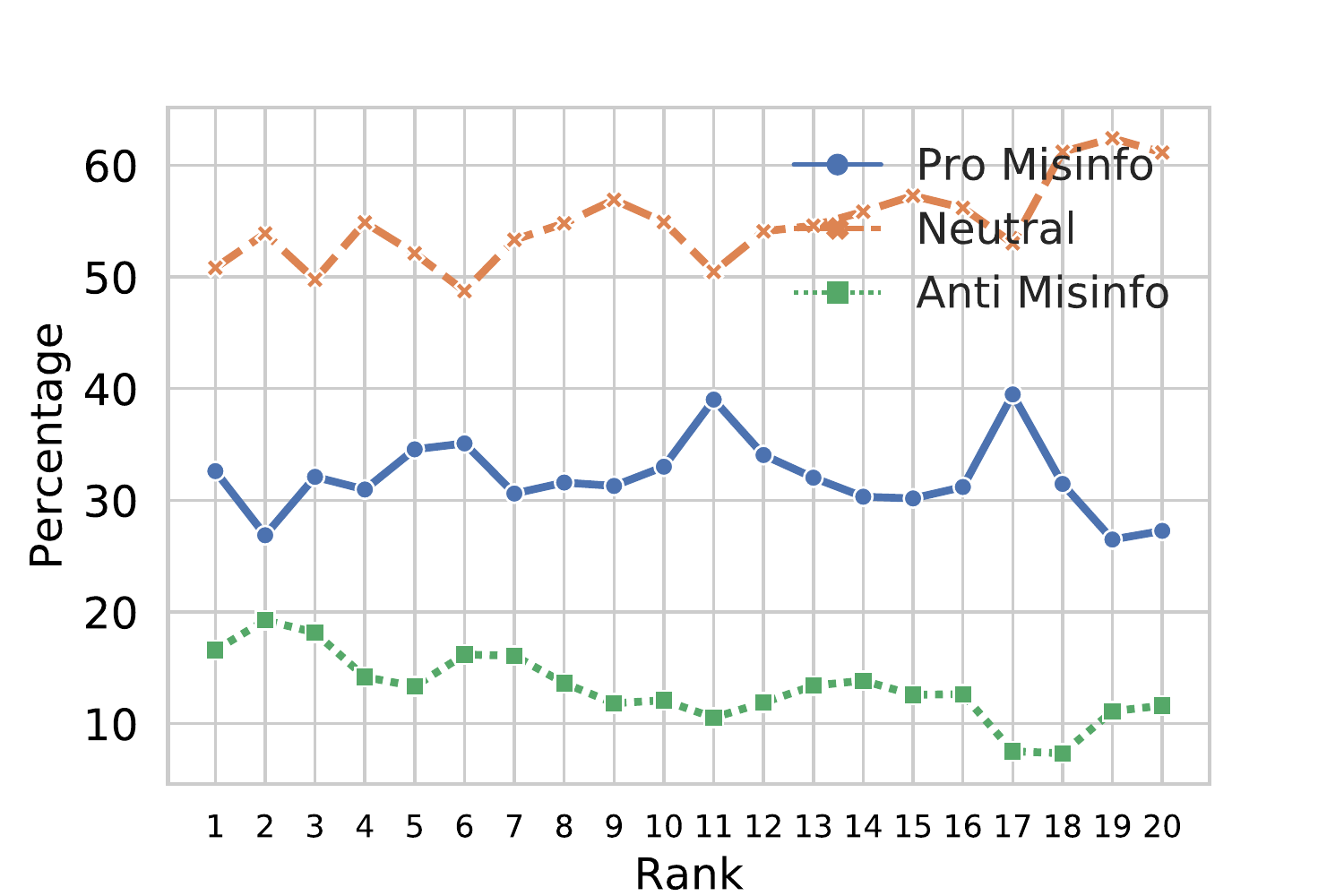} }}\\
  \subfloat[\scriptsize Newest Arrivals]{{ \label{fig:ranking_newest_arrivals_ranks}\includegraphics[scale=0.3]{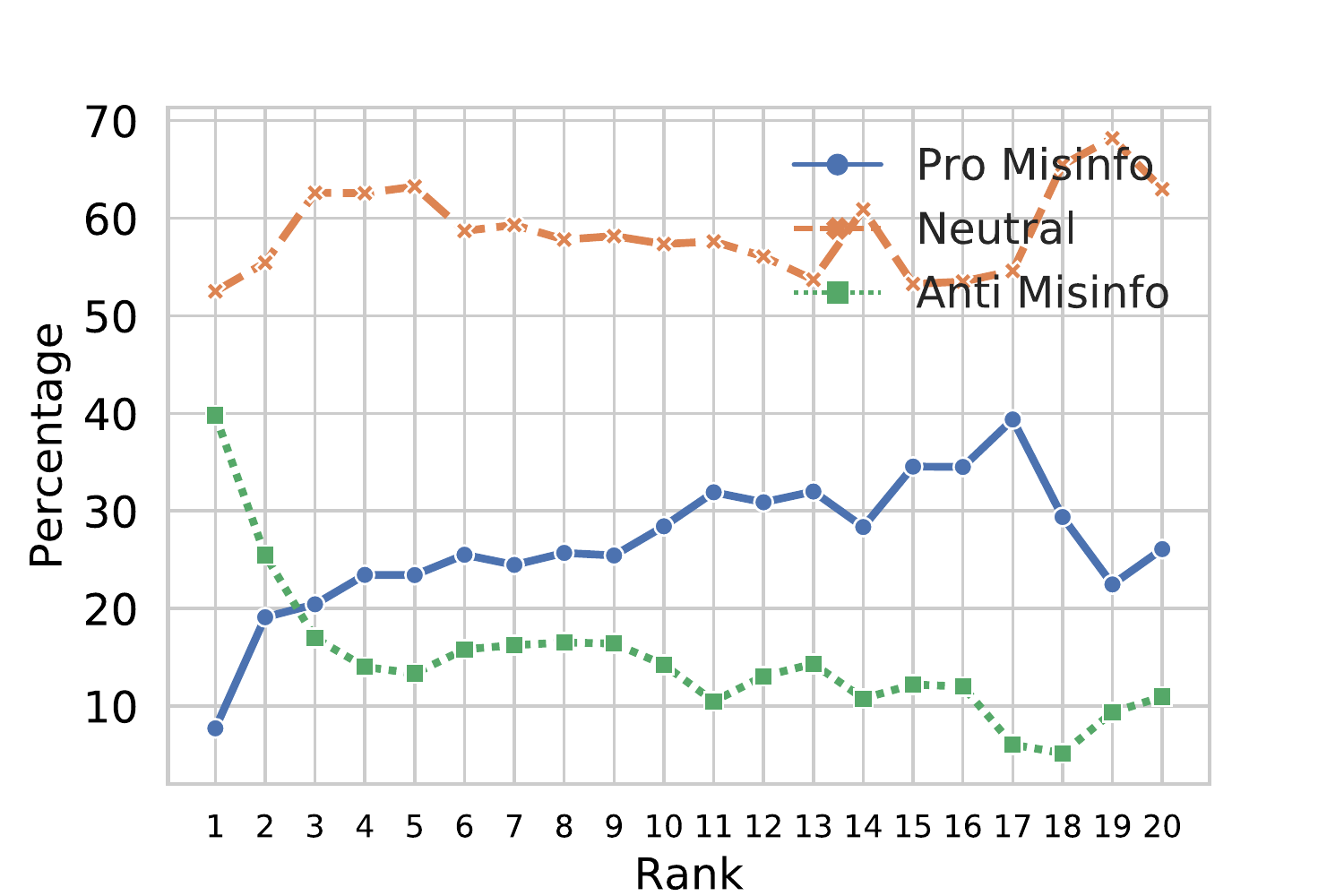} }}
  \vspace{-5pt}
  \caption{\scriptsize Percentages of items ranked at each of the top 20 positions within each annotation category ({\color[HTML]{3b4e6d}Pro Misinfo (1)}, {\color[HTML]{a25325}Neutral (0)} and {\color[HTML]{2f7942}Anti Misinfo (-1)}) for the 5 Amazon search algorithms: (a) Featured algorithm (\emph{\textbf{default}}), (b) Average customer review, (c) Item price ascending, (d) Item price descending and (e) Newest Items added to Amazon}
  \label{fig:rankings}
   \vspace{-15pt}
\end{figure}
\end{scriptsize}

\subsubsection{RQ1b: Query effect}\label{results:RQ1b}
We analyze the effect of the query stance on the stance of search results to understand if there is positive correlation between the query stance and the stance (SERP-MS Score) of the SERP generated by Amazon. We test for a significant difference between SERP-MS of three groups of search queries annotated for either being pro (1), neutral (0) or anti (-1) misinformation. We find that there is a significant difference (\textit{Kruskal Wallis H(2)$\approx$3623.86, p=0.0}) where the means of pro, neutral and anti misinformation queries are \textit{0.26}, \textit{0.128} and \textit{-0.627} respectively. We conclude that the stance of a query positively correlates with the stance of its results, where pro misinformation queries tend to generate more misinformative results than neutral queries and neutral queries generate more misinformative results than anti misinformation queries. Fig. \ref{fig:queries_vs_serps} depicts the frequencies and the variations of SERP-MS scores resulting from each of the three query stances.

%%%% query vs results graphs
\begin{scriptsize}
\begin{figure}[t]
\vspace{-1pt}
  \centering
  \subfloat[]{{\label{fig:queries_vs_serps_frequency}\includegraphics[scale=0.3]{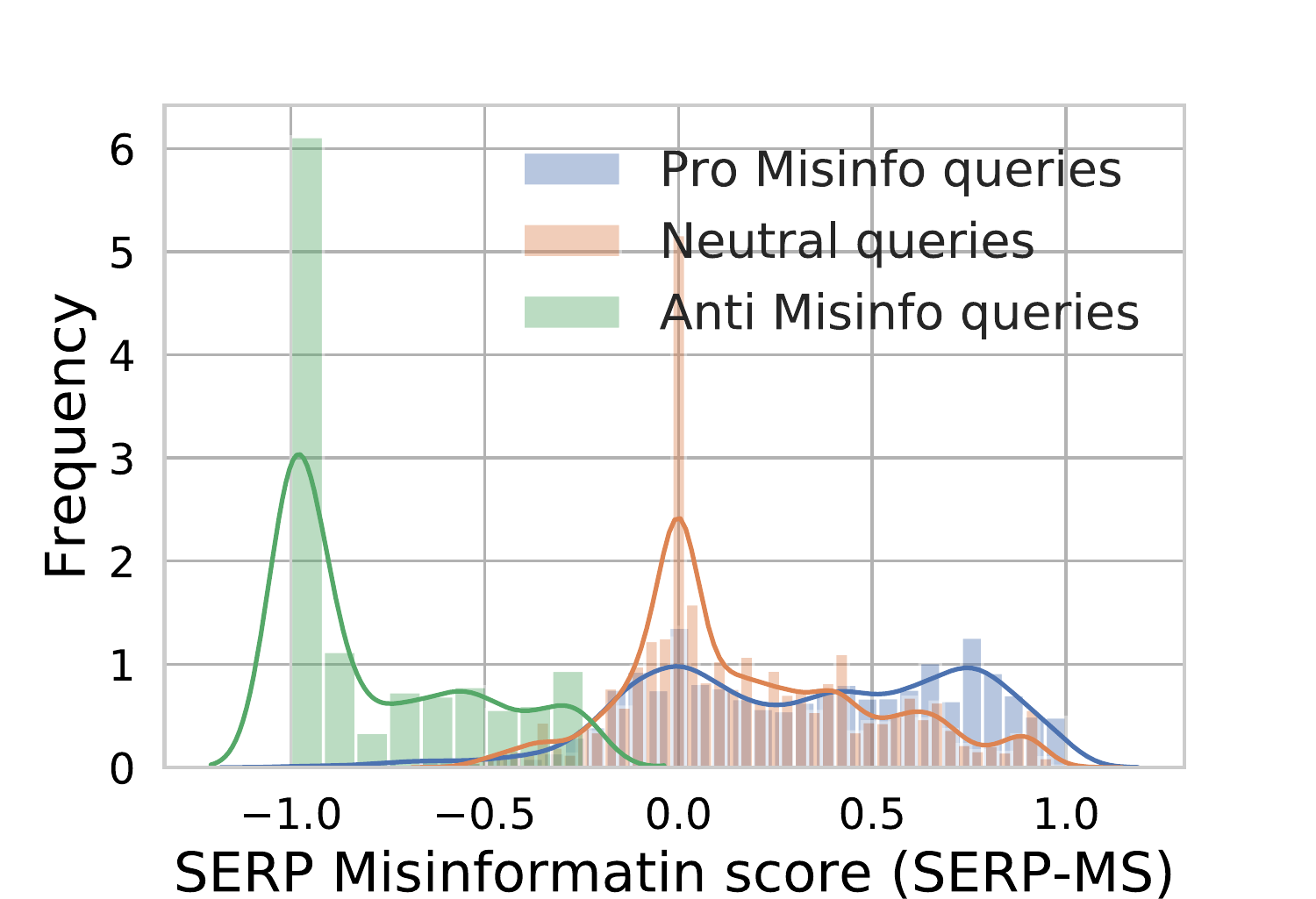} }}
  \subfloat[]{\label{fig:queries_vs_serps_boxplot}{\includegraphics[scale=0.3]{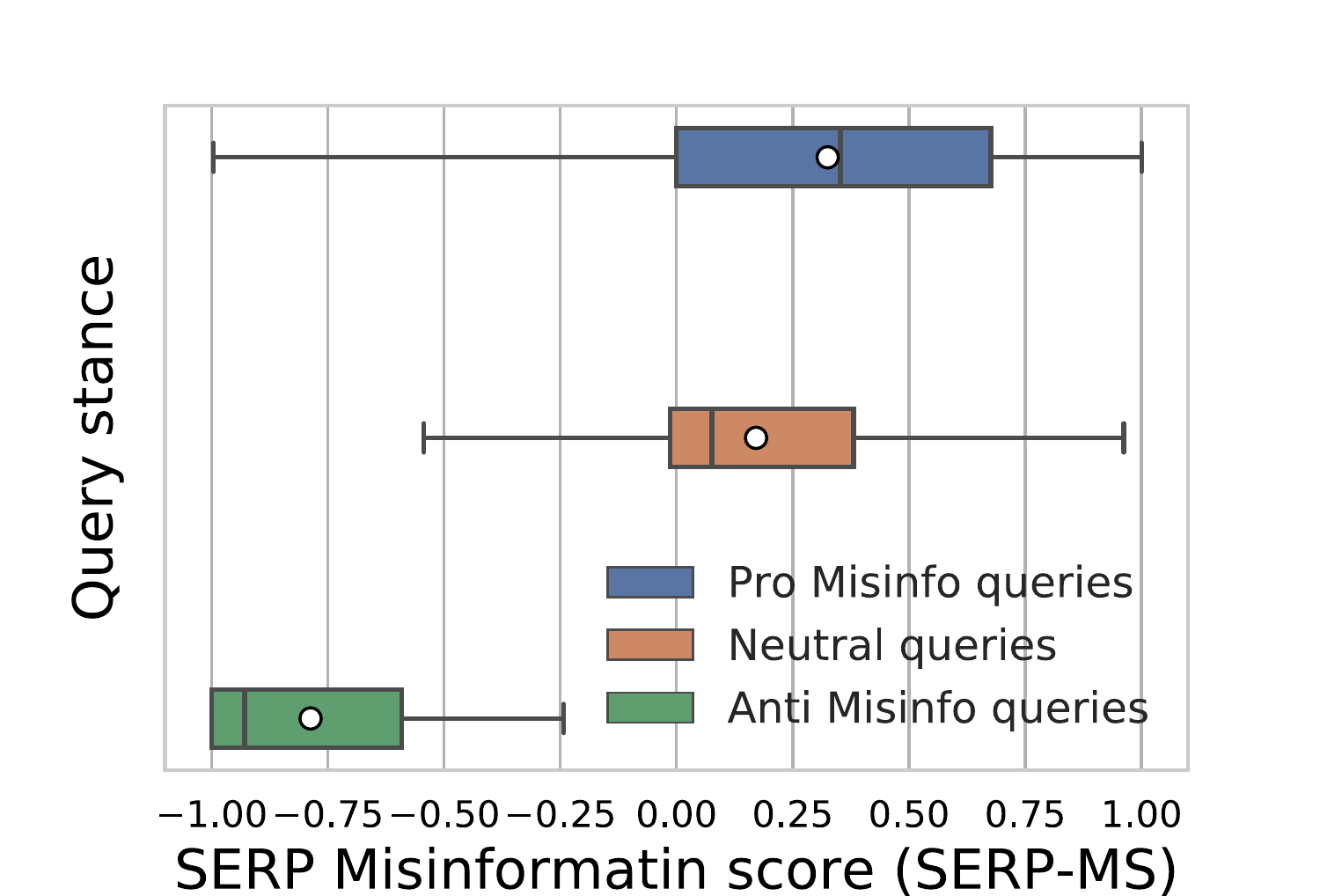}}}
  \vspace{-5pt}
  \caption{\scriptsize (a) Frequency distributions and (b) box plots of the SERP-MS scores for each of the three query stances ({\color[HTML]{3b4e6d}Pro Misinfo (1)}, {\color[HTML]{a25325}Neutral (0)} and {\color[HTML]{2f7942}Anti Misinfo (-1)}).
  }
  \label{fig:queries_vs_serps}
\vspace{-10pt}
\end{figure}
\end{scriptsize}

\subsubsection{RQ1c: User Rating Effect}\label{results:RQ1c}
To investigate the relationship between user rating of an item and its ranking as a search result or a recommendation given its misinformation stance, we depict box plots in fig.~\ref{fig:ratings} of: (a) \textit{ratings} as a measure of item's \textit{preference}, and (b) \textit{number of ratings} as a measure of item's \textit{popularity} of unique search results and recommendations. We find that pro misinformation items have higher ratings than anti misinformation items in both search results and recommendations. On the other hand, neutral items have similar and lower ratings than misinformative items in search results and recommendations respectively. While for items popularity, neutral items are more popular than the other two stances in both search results and recommendations, while misinformative items are more popular than anti-misinformation items in both search results and recommendations.\\
%%%% ratings graphs
\begin{scriptsize}
\begin{figure}[t]
  \centering
  \vspace{-4mm}
  \hspace{-10pt}
  \subfloat[\scriptsize Search Results (Ratings)]{{ \label{fig:serp_ratings}\includegraphics[scale=0.28]{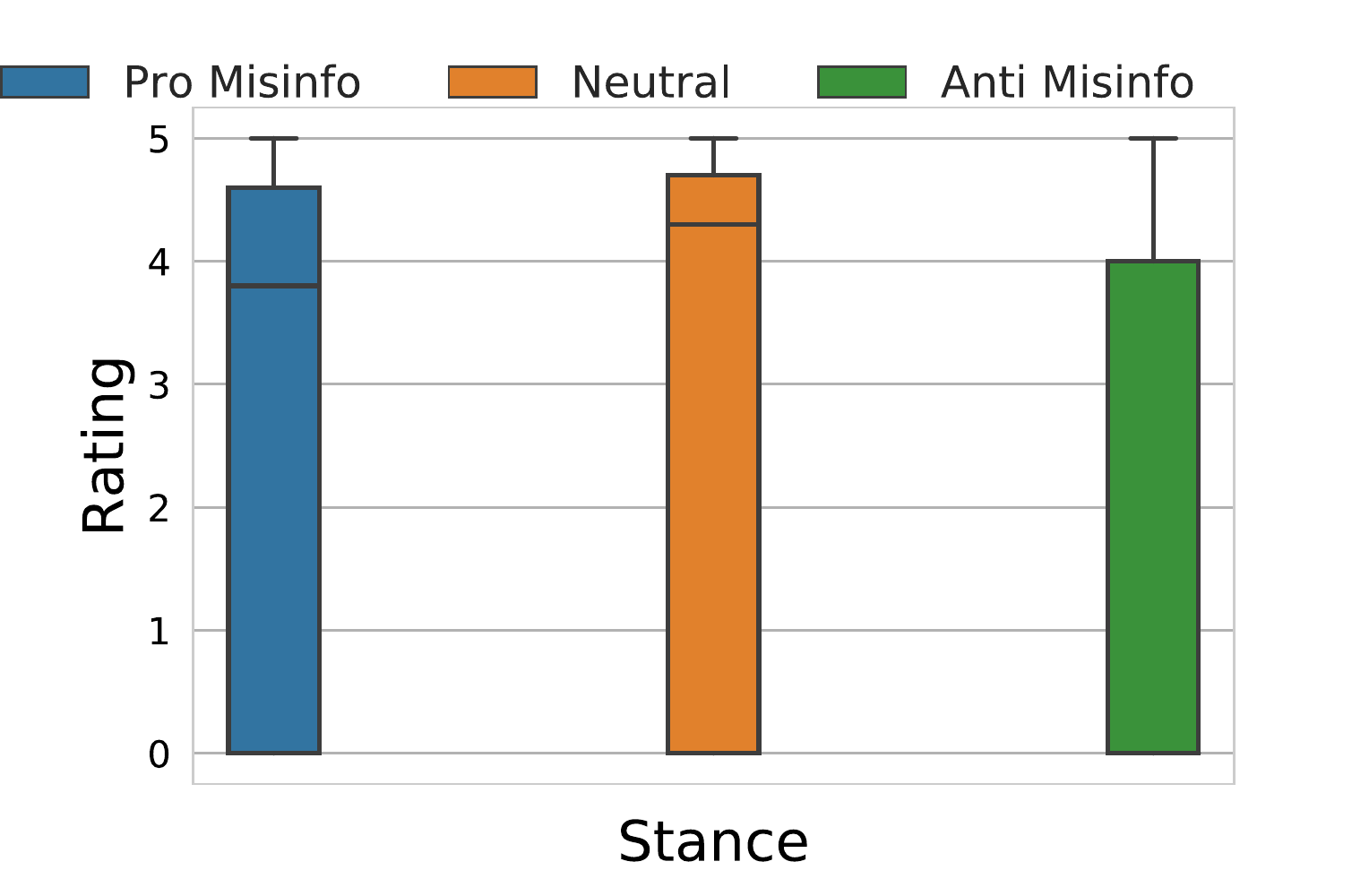} }}
  \subfloat[\scriptsize Search Results (No. of Ratings)]{{ \label{fig:serp_num_ratings}\includegraphics[scale=0.28]{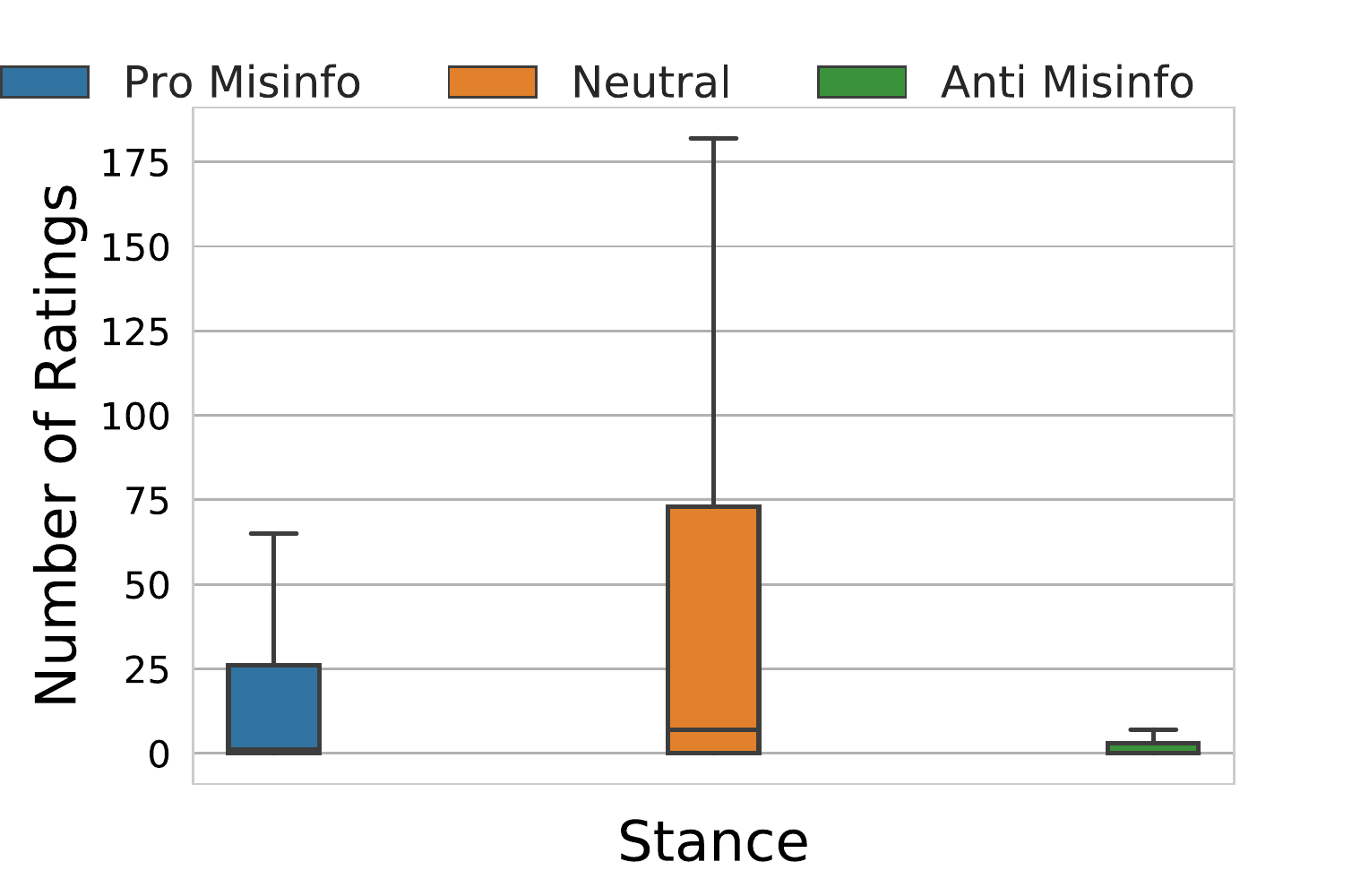} }}\\
  \subfloat[\scriptsize Recommendations (Ratings)]{{ \label{fig:recomm_ratings}\includegraphics[scale=0.28]{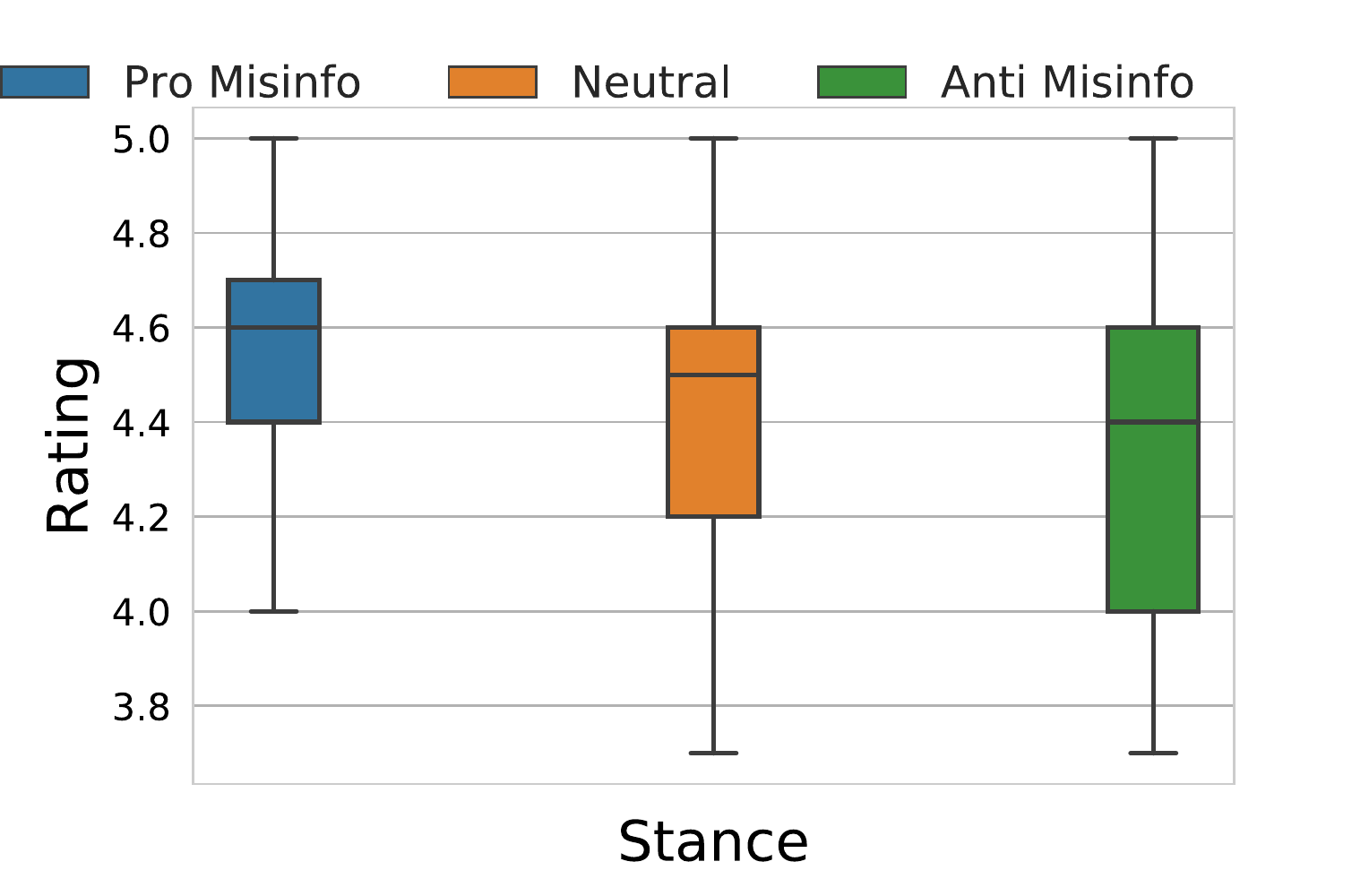} }}
  \subfloat[\scriptsize Recommendations (No. of Ratings)]{{ \label{fig:recomm_num_ratings}\includegraphics[scale=0.28]{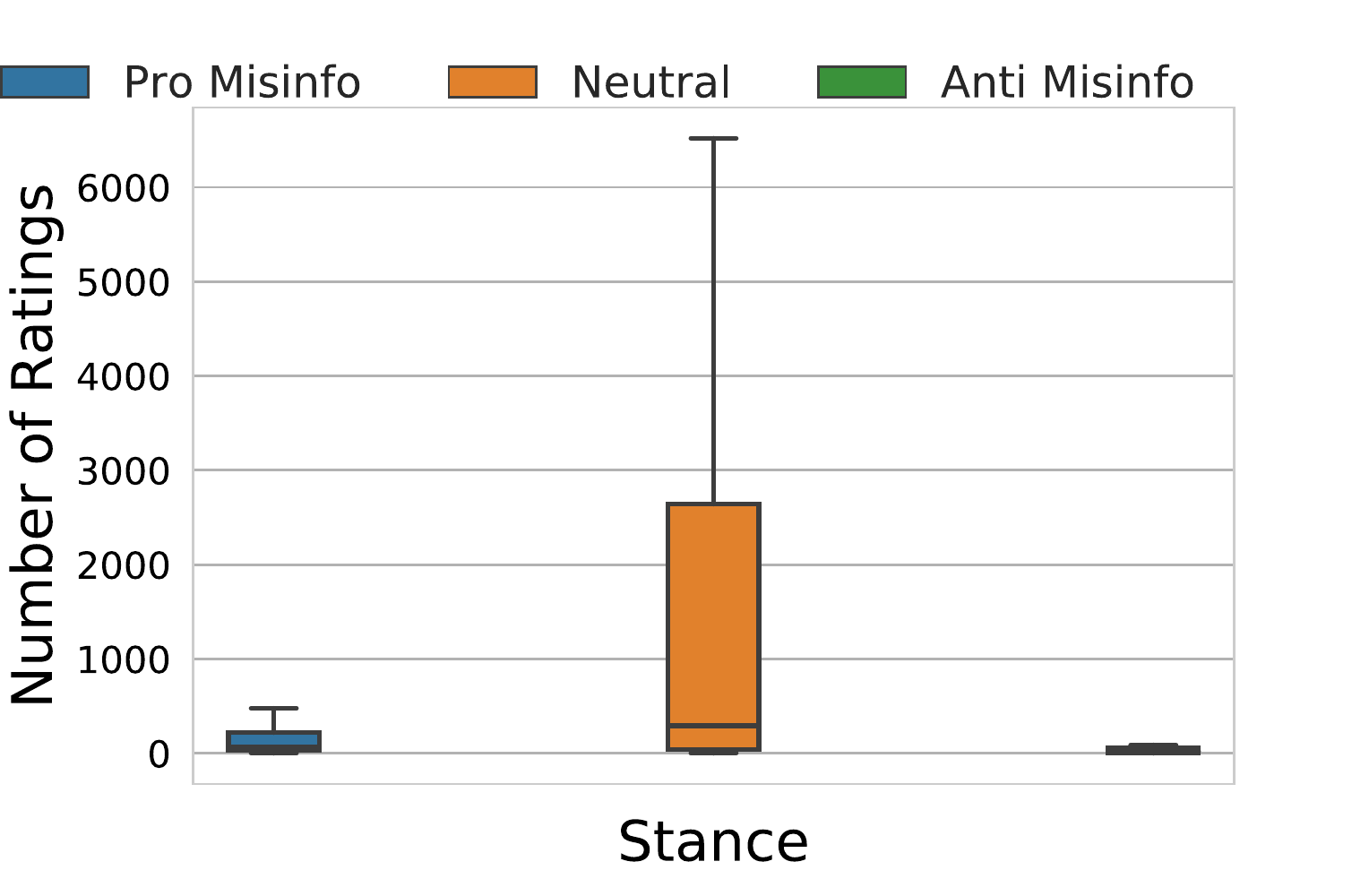} }}
  \caption{\scriptsize Ratings and number of ratings of items annotated as one of the three stances ({\color[HTML]{3b4e6d}Pro Misinfo (1)}, {\color[HTML]{a25325}Neutral (0)} and {\color[HTML]{2f7942}Anti Misinfo (-1)})}
  \label{fig:ratings}
\vspace{-5pt}
\end{figure}
\end{scriptsize}

\vspace{-10pt}
\subsection{RQ2: User Activity Effect}
We investigate the effects of four user activities (search, browse, add to wish list, add to cart) on the misinformation scores SERP-MS and FSERP-MS of SERP pages and homepages recommendations (\emph{After-Search}), respectively. For the collected SERPs, we find no significant difference (\textit{Kruskal Wallis H(3)$\approx$0.66, p$\approx$0.88}) between the SERP-MS scores of accounts that searched after browsing, adding items to wish list or cart and the account that only searched without having prior history. We deduce that performing any of those activities before searching does not affect search results. While for homepage recommendations, we find a significant difference (\textit{Kruskal Wallis H(3)$\approx$25.65, p$\approx$1.13e-5}) between the FSERP-MS scores after performing the four activities. We find that only \emph{searching} without having prior history (i.e browsing, wishing or purchasing) does no affect the homepage recommendations at all, whereas the other three activities affect the homepage recommendations, yet their effects are not significantly different from each other. The mean FSERP-MS scores of \emph{search}, \emph{browse}, \emph{add to list} and \emph{add to cart} are \textit{0.0}, \textit{0.163}, \textit{0.166} and \textit{0.185}, respectively. Figures \ref{fig:activity_effect_SERP} and \ref{fig:activity_effect_homepage} show the frequency distributions of SERP-MS and FSERP-MS scores across all activities.

\begin{scriptsize}
\begin{figure}[t]
  \centering
  \vspace{-4mm}
  \hspace{-10pt}
  \subfloat[\scriptsize SERP]{{\label{fig:activity_effect_SERP}\includegraphics[scale=0.28]{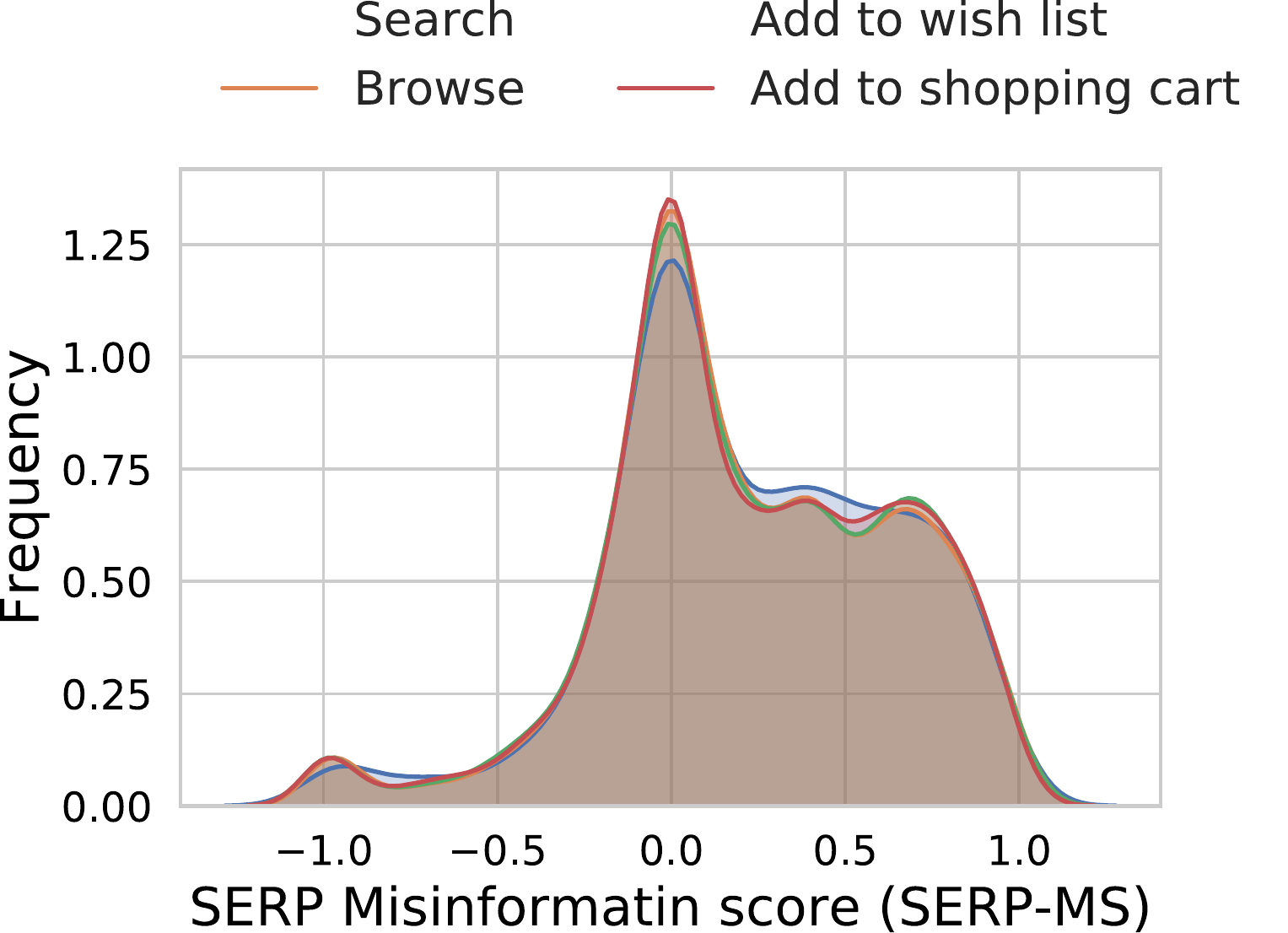} }}
  \subfloat[\scriptsize Homepage]{\label{fig:activity_effect_homepage}{\includegraphics[scale=0.28]{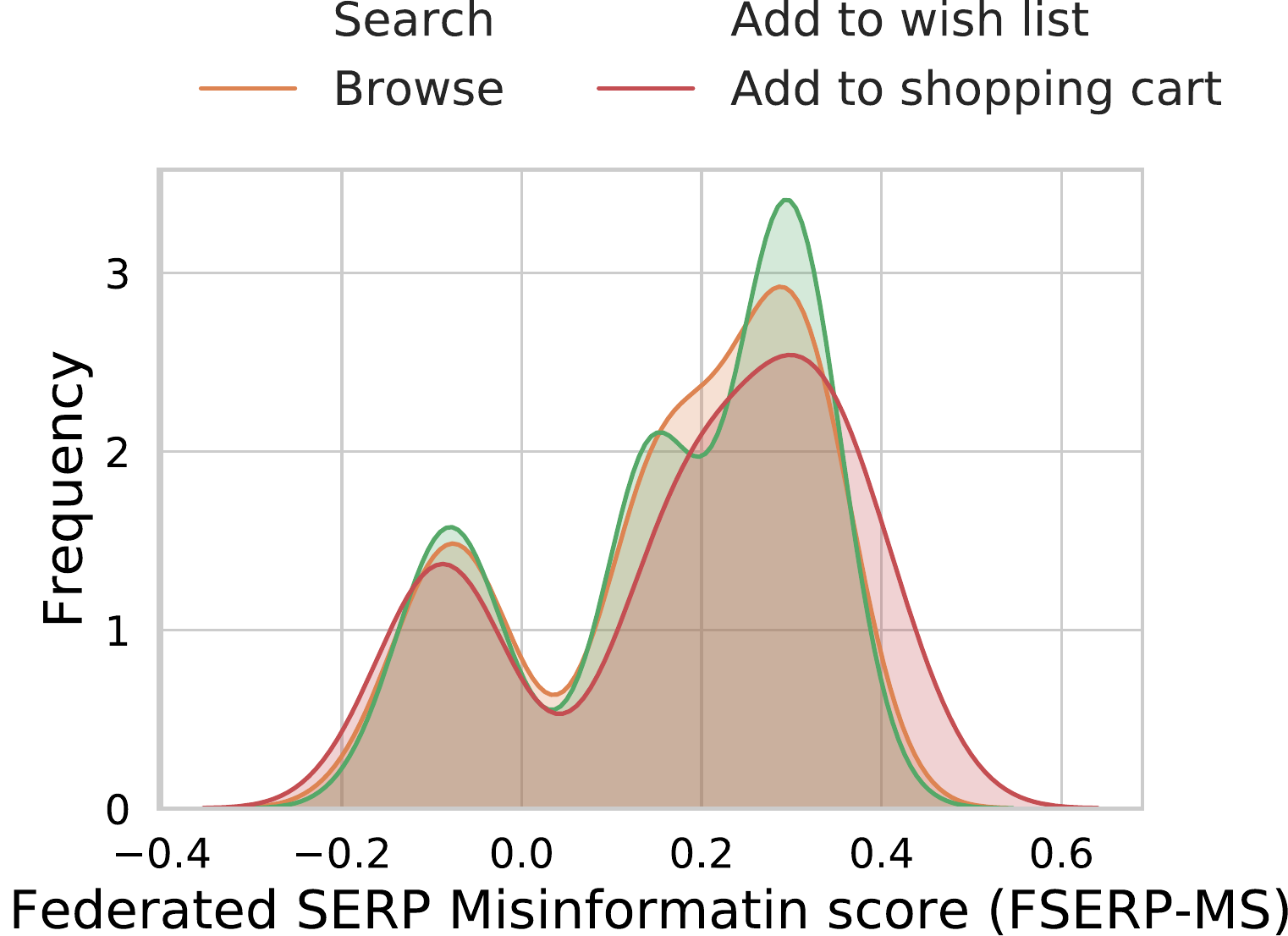}}}
   \vspace{-5pt}
   \centering
   \caption{\scriptsize Frequency distributions of (a) SERP-MS and (b) FSERP-MS scores of SERPs and homepages, respectively, for the four activities (search, browse, add to wish list and add to cart)}
  \label{fig:activity_effect}
\vspace{-10pt}
\end{figure}
\end{scriptsize}

\subsection{RQ3: Item Stance Effect}
We inspect the effect of the four item stance treatments (pro-misinformation, anti-misinformation, neutral and a combination of the three previous treatments) on the calculated misinformation scores SERP-MS and FSERP-MS of the SERP pages and homepages, respectively. We compare the four different treatments SERP-MS scores and find no significant difference (\textit{Kruskal Wallis H(3)$\approx$4.5, p$\approx$0.22}) between the SERP-MS scores of the four treatments regardless of the activity (browsing, wishing or purchasing) done before search. On the other hand, we find a significant difference (\textit{Kruskal Wallis H(3)$\approx$463, p$\approx$5e-100}) between the FSERP-MS scores of homepage recommendations when applying the four treatments of item stance, where accounts that interact (browse, add to wish list or cart) with items that promote misinformation about vaccines get homepage recommendations that also promote more vaccines misinformation than those who do the same activities on the other treatments. We find a positive correlation between the misinformation stance of items that are browsed, added to wish list or shopping cart and the misinformation stance of homepage recommendations after performing any activity. Figures~\ref{fig:treatment_effect_SERP} and \ref{fig:treatment_effect_homepage} show the frequency distributions of SERP-MS and FSERP-MS scores across all stance treatments.

%%%%%%%%%% Effect of treatment
\begin{scriptsize}
\begin{figure}[t]
  \centering
  \vspace{-2mm}
  \hspace{-10pt}
  \subfloat[\scriptsize SERP]{{\label{fig:treatment_effect_SERP}\includegraphics[scale=0.28]{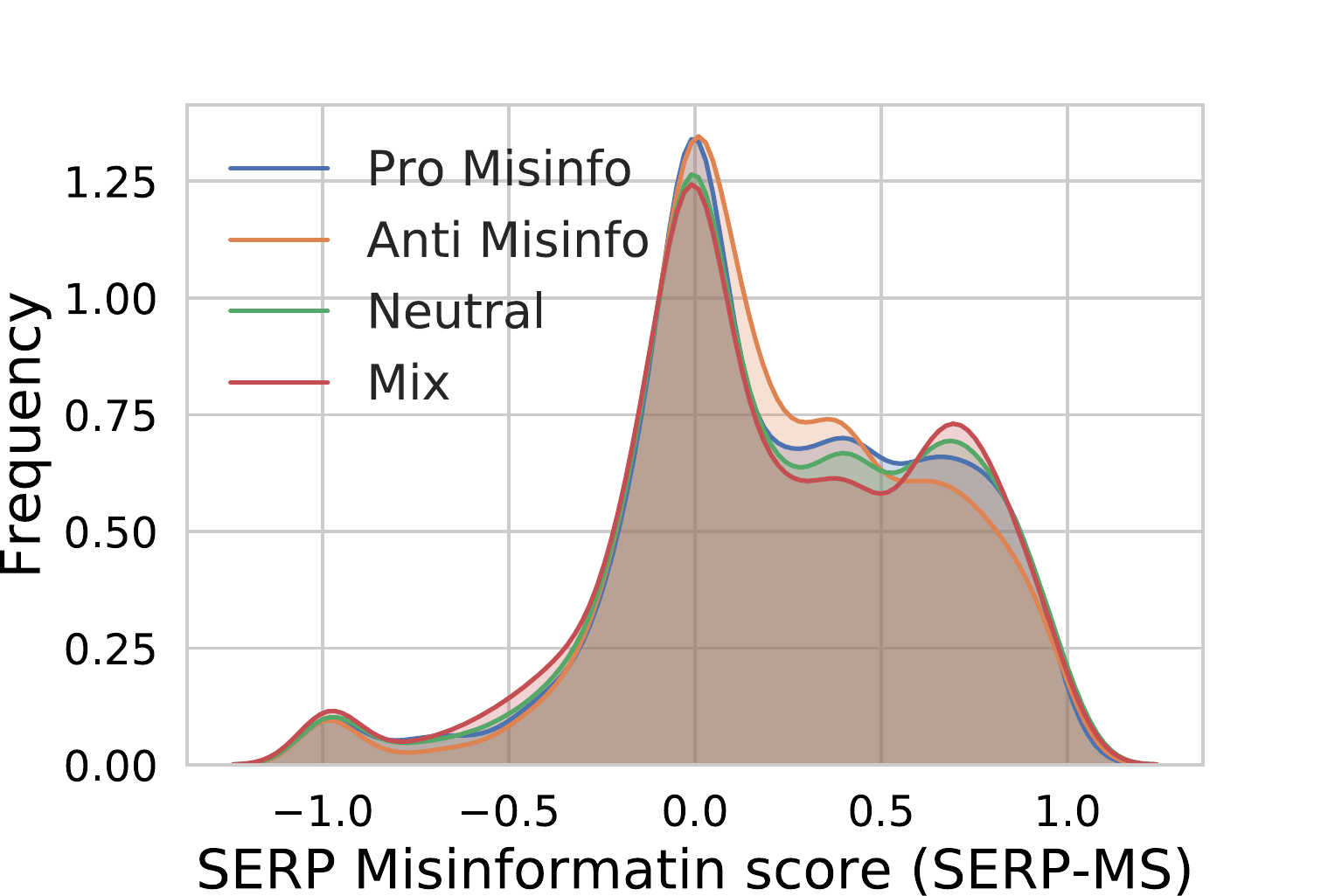} }}
  \subfloat[\scriptsize Homepage]{\label{fig:treatment_effect_homepage}{\includegraphics[scale=0.28]{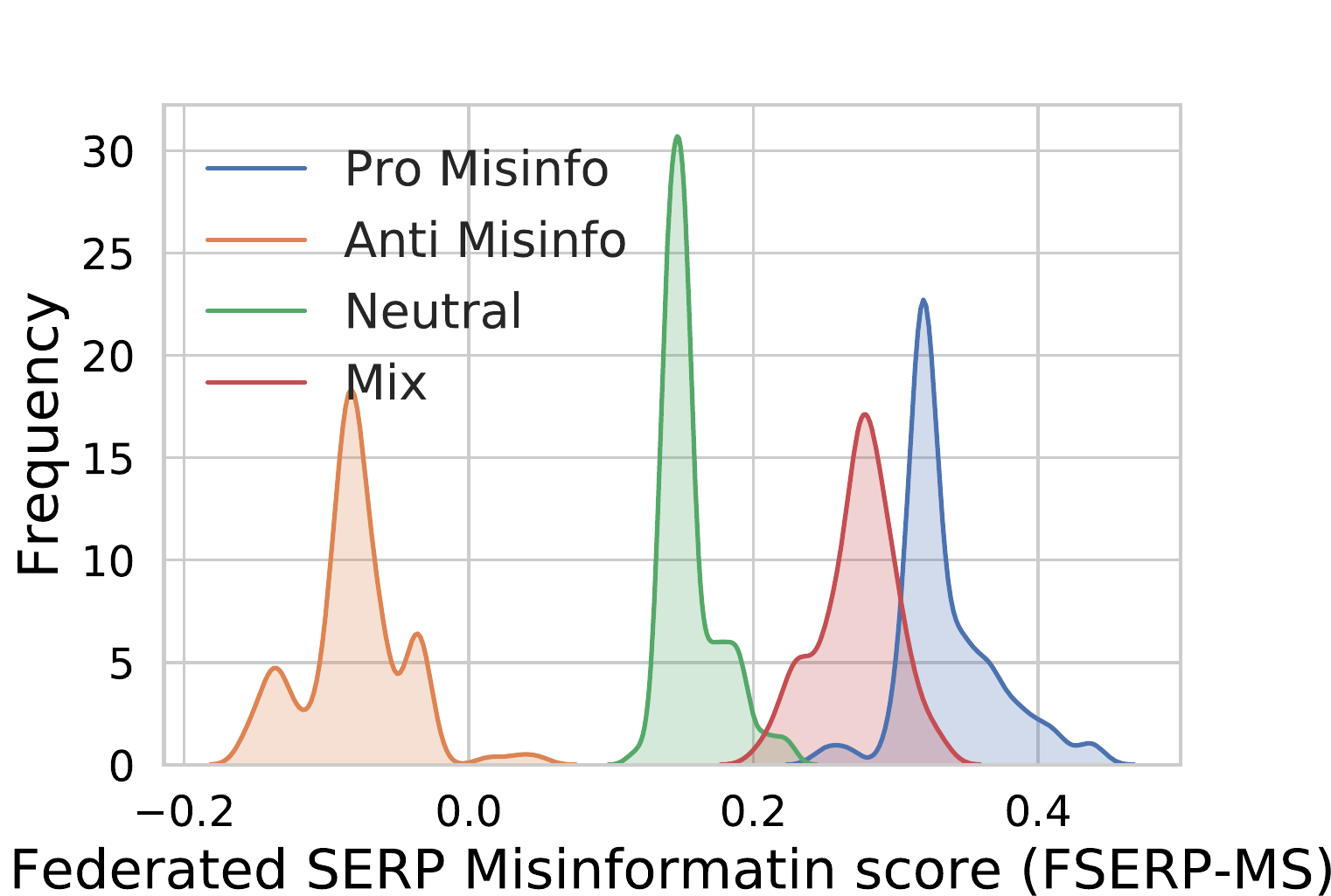}}}
  \vspace{-5pt}
  \caption{\scriptsize Frequency distribution of (a) SERP-MS scores of SERP pages and (b) FSERP-MS scores of homepages for each of the four treatments ({\color[HTML]{3b4e6d}Pro Misinfo (1)}, {\color[HTML]{a25325}Neutral (0)}, {\color[HTML]{2f7942}Anti Misinfo (-1)}, and a combination (mix) of the three other treatments).
  }
  \label{fig:treatment_effect}
   \vspace{-10pt}
\end{figure}
\end{scriptsize}

\section{Discussion}
\textbf{Search Algorithms.} To explore \emph{\textbf{RQ1}}, we analyze three key factors that might affect selecting and ranking items as search results, given that they have stances toward vaccines' misinformation. (1) We compare the five available search algorithms on items rankings, (2) we investigate the effect of the search query on results with respect to their misinformation stance and (3) we inspect users' ratings of items that have misinformation stance. We observe positive correlations between the selection and ranking of items that has a misinformation stance (pro, neutral or anti) and (a) the misinformation stance of search queries (see \ref{results:RQ1b}), (b) the item preference and popularity based on users' ratings (see \ref{results:RQ1c}) and (c) the average customer review (see \ref{results:RQ1a}). We notice that the overall users' evaluation of items (ratings and reviews) plays a critical role in selecting and ranking search results (i.e., the more popular and preferred an item is the more likely it will be highly ranked in search results), also the misinformation stance of a search query positively correlates with the misinformation stance of search results.

\textbf{Effect of personalization based on user activity.} We answer \emph{\textbf{RQ2}} by investigating the effect of 4 main user activities -- search, browse, add to wish list and add to cart --  on the amount of misinformation present to the user in search results and recommendation. We conclude that user's activities with the platform does not affect the misinformation stance of search results. However, having a browsing, wishing or purchasing history with the platform affects the homepage recommendations similarly, in contrast to solely \emph{searching} without having history which has no effect on the homepage recommendations.

\textbf{Effect of personalization based on item's stance.} We answer \emph{\textbf{RQ3}} by investigating the effect when a user interacts with items having misinformation stance -- pro, neutral and anti --  on the amount of misinformation present to the user in search results and recommendations. We find that our hypothesis \emph{H3\textsubscript{a}} is true only for homepage recommendations, which leads to a misinformative filter bubble on a user's homepage. In contrast, the misinformation stance of items being browsed, added to a wish list or a shopping cart does not affect the misinformation stance of search results.

\section{Conclusion}
In this paper, we introduce a methodology for auditing search and recommendation systems for misinformation on online marketplaces. 
Our methodology is the first to inspect how personalization affects the amount of misinformation presented in search results and recommendations on an online marketplace. We applied our methodology and audited the extent of vaccines' misinformative items on Amazon, and the effect of personalization based on (a) user activity (search, browse, add to wish list, add to cart) and (b) misinformation stance of items (pro, neutral and anti) that the user interacts with on the misinformation stance of search results and recommendations. Our  study reveals that the misinformation stance of items selected and ranked by Amazon's search algorithms is positively correlated with both the stance of the search queries and the customers’ evaluation of items (ratings and reviews), (b) search results are not affected by any of the users’ activities or the selection of misinformative items and (c) a misinformative filter bubble forms within the homepage when a user interacts with items of misinformative stance through browsing, adding to a wish list or a shopping cart. In conclusion, we believe that Amazon does not deliberately recommend vaccines' misinformative items, yet mitigating vaccines' misinformation on its marketplace may be an interesting problem to explore.

\begin{spacing}{0.87}
\bibliographystyle{ieeetr}
\bibliography{references}
\end{spacing}

\end{document}